\documentclass[10pt]{article}

\begin{document}

\title{Brans-Dicke Cosmology in $4D$ from scalar-vacuum in $5D$} 
\author{J. Ponce de Leon\thanks{E-mail: jpdel@ltp.upr.clu.edu; jpdel1@hotmail.com}\\ Laboratory of Theoretical Physics, Department of Physics\\ 
University of Puerto Rico, P.O. Box 23343, San Juan, \\ PR 00931, USA} 
\date{March  2010}

\maketitle

\begin{abstract}

We show that Brans-Dicke (BD) theory  in $5D$ may explain the present cosmic accelerated expansion without recurring to  matter fields in $5D$ or dark energy in $4D$. Without making any assumption on the nature of the extra coordinate or the matter content in $5D$, here we demonstrate
that the vacuum BD field equations in $5D$ are equivalent,  on every hypersurface orthogonal to the extra dimension,  to a BD theory in $4D$  with a self interacting potential and an effective matter field. The potential is not introduced by hand, instead the reduction procedure provides an expression that determines its shape up to a constant of integration. It also establishes   the explicit formulae for the effective matter  in $4D$. In the context of FRW cosmologies, we show that the reduced BD theory gives rise to models for accelerated expansion of a matter-dominated universe which are consistent with current observations and with a decelerating radiation-dominated epoch.

\end{abstract}

\medskip



\newpage

\section{Introduction}

One of the most challenging problems in cosmology today is to explain the observed late-time accelerated expansion of the universe \cite{Riess}-\cite{Sievers}. Since the gravity of both baryonic (ordinary) matter and radiation is attractive, an accelerated expansion requires the presence  of a new form of matter (called {\it dark energy}), which could (i) produce gravitational repulsion, i.e., violate the strong energy condition; (ii) account for  $70\%$ of the total content of the universe; (iii) remain unclustered on all scales where gravitational clustering of ordinary matter  is seen. (For a recent review see Ref. \cite{Review}).
Possible candidates for dark energy include: a cosmological constant or a time dependent cosmological term \cite{Peebles}-\cite{Padmanabhan0};  an evolving scalar field known as {\it quintessence} ($Q$-matter) with a potential  giving rise to negative pressure at the present epoch \cite{Zlatev}-\cite{Deustua}; dissipative fluids \cite{DissipativeFluids}; Chaplygin gas \cite{Chaplygin gas 1}-\cite{Chaplygin gas 2}; K-essence \cite{K-essence 1}-\cite{K-essence 4}, and other more exotic models \cite{domain walls}. 

General Relativity (GR) is a well tested theory on solar system scales. In contrast it is poorly tested on cosmic scales \cite{Rachel}. Therefore, the question arises of whether   dark energy  is not just an observational artifact caused by an inappropriate theory of gravity. This and other puzzles of theoretical and experimental gravity have triggered a huge  interest in alternative theories of gravity (See, e.g. \cite{Gilles} and references therein). 
In recent years there has been a renewed interest in scalar-tensor theories of gravity as viable alternatives to general relativity. 
In particular, some researchers \cite{Bertolami}-\cite{Chakraborty} have resorted  to Brans-Dicke theory (BD)  \cite{Dicke} in order to  explain the present accelerated expansion of the universe.  
An attractive feature of BD  
is that the scalar field is a fundamental element of the theory, as opposed to other models in which the scalar field is 
 postulated separately in an ad hoc fashion.
The concept is  that the BD scalar field  could play the role of $Q$-matter or $K$-essence and lead to cosmological acceleration. 
However, it turns out that this is so only in very particular cases: for values of the coupling parameter $\omega$ in the range $- 2 <  \omega < - 3/2$, which not only violate the energy condition on the scalar field but are also inconsistent with a radiation-dominated epoch, unless     $\omega $ varies with time \cite{Pavon}; when there is a
scalar potential, which in turn is  added by hand \cite{Sen}. Further models include the so-called ``chameleon fields" that allow the scalar field to interact with matter (see \cite{Das 2} and references therein). 

Another attempt to derive the present accelerated expansion from a fundamental concept (or postulate) combines the original BD and Kaluza-Klein (KK) theories with the modern view that our $4D$ universe can be  recovered on  a hypersurface orthogonal to the extra dimension. Along these lines,   and assuming  that the extra coordinate  is cyclic (ignorable), it has been shown in \cite{Li} that the BD field equations in $5D$ are equivalent to those of GR in $4D$ with  two scalar fields, viz., the BD scalar field and the $\gamma_{5 5}$ component of the KK metric (For a recent discussion see Ref. \cite{Li 1}). These two scalar fields may account for the present accelerated expansion of the universe, if one admits the existence of matter in $5D$ which does not move along the extra dimension and effectively behaves   as dust in $4D$. 

In this work we adhere to the point of view advanced in \cite{Li}-\cite{Li 1}. However, we propose that neither the cylindricity condition on the extra coordinate nor the higher dimensional  matter hypothesis are necessary. In fact, in this paper we show that, regardless of whether we assume a compact or large extra dimension,  BD theory in $5D$ may explain the present accelerated expansion without introducing matter fields in $5D$ or dark energy in $4D$. We demonstrate that the vacuum BD field equations in $5D$ are equivalent,  on every hypersurface orthogonal to the extra dimension,  to a BD theory in $4D$  with a self interacting potential and an effective matter field. In the context of FRW cosmologies, we show that the reduced BD theory gives rise to models for accelerated expansion of a matter-dominated universe which are consistent with a decelerating radiation-dominated epoch,  for the same value of the BD parameter $\omega$. 

The paper is organized as follows. In section $2$ we perform the dimensional reduction of the scalar-vacuum BD field equations in $5D$.  
Our procedure provides explicit definitions for the effective matter and  potential\footnote{{A similar procedure has recently been discussed in \cite{Aguilar}, where a ``modified" BD theory in $4D$ is obtained from the vacuum BD equations in $5D$. However, our formulae and definitions of the $4D$ quantities  are completely different  from  those derived in \cite{Aguilar}.}} in $4D$.
In section $3$ we study homogeneous and isotropic solutions of the vacuum BD field equations in $5D$ under the assumption of separation of variables (in general, the metric functions may depend on time and the extra coordinate). We show that there are four classes of separable solutions. We discuss in detail the class of power-law solutions, which in turn consists of five families of solutions. 
In section $4$ we study the $4D$ cosmological scenarios allowed by the $5D$ power-law  solutions derived in section $3$. We find  that distinct families of solutions in $5D$ lead to 	very much alike scenarios in $4D$; they can give accelerating matter-dominated era as well as decelerating radiation-dominated era  for the same value of $\omega$ in the range $- 3/2 < \omega < - 1$. In section $5$ we give a summary of our results. Finally, in the Appendix   we show, by means of explicit integration, that the assumption of separability is consistent with the 
field equations.

\section{Dimensional reduction of  Brans-Dicke theory in $5D$}

The Brans-Dicke (BD) theory of gravity in $5D$ is described by the action \cite{Li}
\begin{equation}
\label{action}
S_{(5)} = \int{d^5 x \sqrt{|\gamma^{(5)}|}\left[\phi R^{(5)} - \frac{\omega}{\phi}\gamma^{A B}\left(\nabla_{A}\phi\right)\left(\nabla_{B}\phi\right)  \right]}  + 16 \pi \int{d^5 x \sqrt{|\gamma^{(5)}|}L_{m}^{(5)}}, 
\end{equation}
where $R^{(5)}$ is the curvature scalar associated with the $5D$ metric $\gamma_{A B}$; $\gamma^{(5)}$ is the determinant of $\gamma_{A B}$; $\phi$ is a scalar field; $\omega $ is a dimensionless coupling constant; $L_{m}^{(5)}$ represents the Lagrangian of the matter fields in $5D$ and does not depend on $\phi$.  

The equations for the gravitational field in $5D$ derived from (\ref{action}) read 

\begin{equation}
\label{equation for the metric in 5D}
G_{AB}^{(5)} = R_{AB}^{(5)} - \frac{1}{2}\gamma_{AB}R^{(5)} = \frac{8 \pi}{\phi}T_{AB}^{(5)} + \frac{\omega}{\phi^2}\left[\left(\nabla_{A}\phi\right) \left(\nabla_{B}\phi\right) - \frac{1}{2}\gamma_{AB} \left(\nabla^{C}\phi\right)\left(\nabla_{C}\phi\right)\right] + \frac{1}{\phi}\left(\nabla_{A}\nabla_{B}\phi - \gamma_{AB} \nabla^2\phi\right),
\end{equation}
where $\nabla^2 \equiv \nabla_{A}\nabla^{A}$ and $T_{AB}^{(5)}$ represents the energy-momentum tensor (EMT) of matter fields in $5D$ with trace $T^{(5)} = \gamma^{AB} T_{AB}^{(5)}$. For the sake of generality of the reduction procedure in this section $T_{AB}^{(5)} \neq 0$.

The field equation for the scalar field $\phi$ is determined by (\ref{action}) as
\begin{equation}
\label{dalambertian in 5D in terms of R5}
\frac{2 \omega}{\phi}\nabla^2 \phi - \frac{\omega}{\phi^2}\left(\nabla_{A}\phi\right)\left(\nabla^A \phi\right) + R^{(5)} = 0.
\end{equation}
Taking the trace of (\ref{equation for the metric in 5D}) we find 
\begin{equation}
\label{R5}
R^{(5)} = - \frac{16 \pi}{3 \phi} T^{(5)} + \frac{\omega}{\phi^2}\left(\nabla_{A}\phi\right)\left(\nabla^{A}\phi\right) + \frac{8}{3 \phi}\nabla^2 \phi.
\end{equation}
Combining the last two equations we get 
\begin{equation}
\label{equation for phi}
\nabla^2\phi = \frac{8 \pi}{4 + 3\omega} T^{(5)}.
\end{equation}

\medskip

In this work we  use coordinates where the metric in $5D$ can be written as\footnote{Notation:  $x^{\mu} = (x^0, x^1, x^2, x^3)$ are the coordinates in $4D$ and $y$ is the coordinate along the extra dimension.  We use spacetime signature $(+, -, -, -)$, while $\epsilon = \pm 1$ allows for spacelike or timelike extra dimension. }  
\begin{equation}
\label{line element in 5D}
dS^2 = \gamma_{AB}d x^A d x^B = g_{\mu\nu}(x, y)d x^{\mu}d x^{\nu} + \epsilon \Phi^2(x, y) d y^2,
\end{equation}
in such a way that  our $4D$ spacetime can be recovered by going onto a hypersurface $\Sigma_{y}:y =  y_{0} = $ constant, which is orthogonal to the $5D$ unit vector
\begin{equation}
\label{unit vector n}
{\hat{n}}^{A} = \frac{\delta^{A}_{4}}{\Phi}, \;\;\;n_{A}n^{A} = \epsilon,
\end{equation}
along the extra dimension, and $g_{\mu\nu}$ can be interpreted as the metric of the spacetime.

The effective field equations (FE) in $4D$ are obtained from dimensional reduction of  (\ref{equation for the metric in 5D}) and (\ref{equation for phi}). 
To achieve such a reduction we note that 
\begin{eqnarray}
\nabla_{\mu}\nabla_{\nu}\phi &=& D_{\mu}D_{\nu}\phi + \frac{\epsilon}{2 \Phi^2}\stackrel{\ast}g_{\mu\nu}\stackrel{\ast}\phi
, \nonumber \\
\nabla_{4}\nabla_{4}\phi &=&  \epsilon \Phi \left(D_{\alpha}\Phi\right)\left(D^{\alpha}\phi\right) + \stackrel{\ast \ast}\phi -  \frac{\stackrel{\ast }\Phi}{\Phi}\stackrel{\ast }\phi,\nonumber \\
 \nabla^2\phi &=& D^2 \phi + \frac{\left(D_{\alpha}\Phi\right)\left(D^{\alpha}\phi\right)}{\Phi} + \frac{\epsilon}{\Phi^2}\left[\stackrel{\ast \ast}\phi + \stackrel{\ast}\phi \left(\frac{g^{\mu\nu}\stackrel{\ast}g_{\mu\nu}}{2} - \frac{\stackrel{\ast}\Phi}{\Phi}\right)\right] 
\end{eqnarray}
where the asterisk denotes partial derivative with respect to the extra coordinate (i.e., $\partial/\partial y = \ast$);  $D_{\alpha}$ is the covariant derivative on $\Sigma_{y}$, which is calculated with $g_{\mu\nu}$, and $D^2 \equiv D^{\alpha}D_{\alpha}$. 

Using these expressions, the spacetime components ($A = \mu, B = \nu$) of the $5D$ field equations (\ref{equation for the metric in 5D}) can be written as 

\begin{eqnarray}
\label{spacetime components of G5}
G_{\mu\nu}^{(5)} = && \frac{8 \pi}{\phi}T_{\mu\nu}^{(5)} + \frac{\omega}{\phi^2}\left[\left(D_{\mu}\phi\right) \left(D_{\nu}\phi\right) - \frac{1}{2}g_{\mu\nu} \left(D_{\alpha}\phi\right)\left(D^{\alpha}\phi\right)\right] + \frac{1}{\phi}\left(D_{\mu}D_{\nu}\phi - g_{\mu\nu} D^2\phi\right)\nonumber \\
&& - \frac{g_{\mu\nu}\left(D_{\alpha}\Phi\right)\left(D^{\alpha}\phi\right)}{\Phi \phi} - \frac{\epsilon g_{\mu\nu}}{2 \Phi^2 \phi}\left[ 2 \stackrel{\ast \ast}\phi +  \stackrel{\ast}\phi \left(g^{\alpha\beta}\stackrel{\ast}g_{\alpha\beta} - 2 \frac{\stackrel{\ast}\Phi}{\Phi} + \omega \frac{\stackrel{\ast}\phi}{\phi}\right)\right] + \frac{\epsilon \stackrel{\ast}g_{\mu\nu}\stackrel{\ast}\phi}{2 \Phi^2 \phi}. 
\end{eqnarray}

To construct the Einstein tensor in $4D$ we have to express $R_{\alpha\beta}^{(5)}$ and $R^{(5)}$ in terms of the corresponding $4D$ quantities.
The Ricci tensor $R^{(4)}_{\mu\nu}$ of the metric $g_{\mu\nu}$ and the scalar field $\Phi$ are related to the Ricci tensor $R_{AB}^{(5)}$ of $\gamma_{AB}$ by \cite{WJPdeL}
\begin{eqnarray}
\label{dimensional reduction}
R_{\alpha\beta}^{(5)} &=& R_{\alpha\beta}^{(4)} - \frac{D_{\alpha}D_{\beta}\Phi}{\Phi} + \frac{\epsilon}{2\Phi^2}\left(\frac{\stackrel{\ast}\Phi \stackrel{\ast}g_{\alpha\beta}}{\Phi} - \stackrel{\ast \ast}g_{\alpha\beta} + g^{\lambda\mu}\stackrel{\ast}g_{\alpha\lambda} \stackrel{\ast} g_{\beta\mu} - \frac{g^{\mu\nu}\stackrel{\ast }g_{\mu\nu}\stackrel{\ast}g_{\alpha\beta}}{2}\right), \nonumber \\
R_{44}^{(5)} &=& - \epsilon \Phi D^2 \Phi - \frac{\stackrel{\ast}g^{\lambda\beta}\stackrel{\ast}g_{\lambda\beta}}{4} - \frac{g^{\lambda\beta}\stackrel{\ast \ast}g_{\lambda\beta}}{2} + \frac{\stackrel{\ast}\Phi g^{\lambda\beta}\stackrel{\ast}g_{\lambda\beta}}{2\Phi}.
\end{eqnarray}
From (\ref{equation for the metric in 5D})-(\ref{R5}) and the second equation in (\ref{dimensional reduction}) we obtain 
\begin{eqnarray}
\label{equation for Phi}
\frac{D^2 \Phi}{\Phi}  = 
 &-& \frac{\left(D_{\alpha}\Phi\right)\left(D^{\alpha}\phi\right)}{\Phi \phi} 
- \frac{\epsilon}{2 \Phi^2}\left[{g^{\lambda\beta}\stackrel{\ast \ast}g_{\lambda\beta}}  + \frac{\stackrel{\ast}g^{\lambda\beta}\stackrel{\ast}g_{\lambda\beta}}{2} -  \frac{\stackrel{\ast}\Phi g^{\lambda\beta}\stackrel{\ast}g_{\lambda\beta}}{\Phi}\right] - \frac{\epsilon}{\Phi^2 \phi}\left[\stackrel{\ast \ast}\phi + \stackrel{\ast}\phi\left(\frac{\omega \stackrel{\ast}\phi}{\phi} - \frac{\stackrel{\ast}\Phi}{\Phi}\right)\right]\nonumber 
\\ &+& \frac{8\pi}{\phi}\left[\frac{\left(\omega + 1\right) T^{(5)}}{4 + 3\omega} - \frac{\epsilon T_{44}^{(5)}}{\Phi^2}\right].
\end{eqnarray}
Substituting this expression into $R^{(5)} = \gamma^{AB}R_{AB}$ we find 
\begin{eqnarray}
\label{R5 in terms of R4}
R^{(5)} =  R^{(4)}
   &+& \frac{2 \left(D_{\alpha}\Phi\right)\left(D^{\alpha}\phi\right)}{\Phi \phi}\nonumber  - \frac{\epsilon}{4\Phi^2}\left[\stackrel{\ast}g^{\alpha\beta}\stackrel{\ast}g_{\alpha\beta} + \left(g^{\alpha\beta}\stackrel{\ast}g_{\alpha\beta}\right)^2\right]
+ \frac{2 \epsilon}{\Phi^2 \phi}\left[ \stackrel{\ast \ast}\phi\ + \stackrel{\ast}\phi\left(\frac{\omega \stackrel{\ast}\phi}{\phi} - \frac{ \stackrel{\ast}\Phi}{\Phi}\right) \right]\\
 &+& \frac{16 \pi}{\phi}\left[\frac{\epsilon T_{44}^{(5)}}{\Phi^2} - \frac{\left(\omega + 1\right) T^{(5)}}{4 + 3 \omega}\right],
\end{eqnarray}
where $R^{(4)} = g^{\alpha\beta}R_{\alpha\beta}^{(4)}$ is the scalar curvature of the spacetime hypersurfaces $\Sigma_{y}$.

\medskip

$\bullet$ We are now ready to obtain the effective equations for gravity in $4D$. With this aim we substitute the first equation in (\ref{dimensional reduction}) and (\ref{R5 in terms of R4}) into (\ref{spacetime components of G5}) and isolate $G_{\mu\nu}^{(4)} = R_{\mu\nu}^{(4)} - g_{\mu\nu} R^{(4)}/2$. The result can be written as
\begin{eqnarray}
\label{field equations for g}
G_{\mu\nu}^{(4)} =  \frac{8\pi}{\phi}\left(S_{\mu\nu} + T_{\mu\nu}^{(BD)}\right) 
 + \frac{\omega}{\phi^2}\left[\left(D_{\mu}\phi\right) \left(D_{\nu}\phi\right) - \frac{1}{2}g_{\mu\nu} \left(D_{\alpha}\phi\right)\left(D^{\alpha}\phi\right)\right] + \frac{1}{\phi}\left(D_{\mu}D_{\nu}\phi - g_{\mu\nu} D^2\phi\right) - g_{\mu\nu}\frac{V(\phi)}{2 \phi}
\end{eqnarray}
where we have introduced the quantity  $V(\phi)$, which (as we will see bellow) plays the role of an effective or induced scalar potential; $S_{\mu\nu}$ is the reduced EMT of the matter fields in $5D$
\begin{equation}
\label{definition of S}
S_{\mu\nu} = T_{\mu\nu}^{(5)} - g_{\mu\nu}\left[\frac{\left(\omega + 1\right) T^{(5)}}{4 + 3 \omega} - \frac{\epsilon T_{44}^{(5)}}{\Phi^2}\right],
\end{equation}
and $T_{\mu\nu}^{(BD)}$ can be interpreted as an induced  EMT for an effective BD theory in $4D$. It is given by

\begin{equation}
\label{T(BD)}
8\pi T_{\mu\nu}^{(BD)} = 8 \pi T_{\mu\nu}^{(STM)} + \frac{\epsilon \stackrel{\ast}\phi}{2 \Phi^2 }\left[\stackrel{\ast}g_{\mu\nu} + g_{\mu\nu}\left(\frac{\omega \stackrel{\ast}\phi}{\phi} - g^{\alpha\beta}\stackrel{\ast}g_{\alpha\beta}\right)\right] + \frac{1}{2}g_{\mu\nu}V,
\end{equation}
with 
\begin{eqnarray}
\label{T(IMT)}
\frac{8\pi}{\phi}T_{\mu\nu}^{(STM)} &\equiv& \frac{D_{\mu}D_{\nu}\Phi}{\Phi} - \frac{\epsilon}{2\Phi^2}\left\{\frac{\stackrel{\ast}\Phi \stackrel{\ast}g_{\mu\nu}}{\Phi} - \stackrel{\ast \ast}g_{\mu\nu} + g^{\alpha\beta}\stackrel{\ast}g_{\mu\alpha}\stackrel{\ast}g_{\nu\beta} - \frac{g^{\alpha\beta}\stackrel{\ast}g_{\alpha\beta}\stackrel{\ast}g_{\mu\nu}}{2} + \frac{g_{\mu\nu}}{4}\left[\stackrel{\ast}g^{\alpha\beta}\stackrel{\ast}g_{\alpha\beta} + \left(g^{\alpha\beta}\stackrel{\ast}g_{\alpha\beta}\right)^2\right]\right\}.
\end{eqnarray}
Since in BD $\phi$ acts as the inverse of the Newtonian gravitational constant $G$, (\ref{T(IMT)}) is identical to the induced EMT used in STM (Space-Time-Matter theory)   \cite{WJPdeL}.  The second term in (\ref{T(BD)}) depends on the first derivatives of $\phi$ with respect to the fifth coordinate and represents the  effective EMT in $4D$ coming from the scalar field. The equation for the scalar potential $V$ is given bellow by (\ref{residual term}).

Taking the trace of  (\ref{field equations for g}) we   obtain a simple relation between  $R^{(4)}$, $S = g^{\mu\nu}S_{\mu\nu}$ and  $T^{(BD)} = g^{\mu\nu} T_{\mu\nu}^{(BD)}$, namely  
  (we note that $g^{\mu\nu}T_{\mu\nu}^{(5)} = T^{(5)} - \epsilon T_{44}^{(4)}/\Phi^2$)  
\begin{eqnarray}
\label{R4 in terms of TBD}
R^{(4)} = - \frac{8\pi}{\phi}\left(S +  T^{(BD)}\right) +  \frac{\omega \left(D_{\alpha}\phi\right)\left(D^{\alpha}\phi\right)}{\phi^2} + \frac{3 D^2 \phi}{\phi} + \frac{2 V}{\phi}.
\end{eqnarray}

$\bullet$ To construct the $4D$ counterpart of (\ref{equation for phi}) we substitute   (\ref{R5 in terms of R4}) and (\ref{R4 in terms of TBD}) into (\ref{dalambertian in 5D in terms of R5}). After some manipulations we get    

\begin{equation}
\label{D2 phi with residual term}
D^2\phi = \frac{8\pi}{3 + 2\omega} \left(S +  T^{(BD)}\right) + \frac{1}{3 + 2 \omega}\left[\phi \frac{d V(\phi)}{d \phi} - 2 V(\phi)\right],
\end{equation}
where 
\begin{eqnarray}
\label{residual term}
\phi \frac{dV(\phi)}{d \phi} \equiv  &-& {2\left(1 + \omega\right)}\left[\frac{\left(D_{\alpha}\Phi\right)\left(D^{\alpha}\phi\right)}{\Phi} + \frac{\epsilon}{\Phi^2}\left(\stackrel{\ast \ast}\phi - \frac{\stackrel{\ast}\Phi \stackrel{\ast}\phi}{\Phi}\right)\right]
- \frac{\epsilon \omega \stackrel{\ast}\phi}{\Phi^2} \left(\frac{\stackrel{\ast }\phi}{\phi} + g^{\mu\nu}\stackrel{\ast}g_{\mu\nu}\right)\nonumber \\ 
&+& \frac{\epsilon \phi}{4\Phi^2}\left[\stackrel{\ast}g^{\alpha\beta}\stackrel{\ast}g_{\alpha\beta} + \left(g^{\alpha\beta}\stackrel{\ast}g_{\alpha\beta}\right)^2\right] 
 + {16\pi}\left[\frac{\left(1 + \omega\right)T^{(5)}}{4 + 3\omega} - \frac{\epsilon T_{44}^{(5)}}{\Phi^2}\right].
\end{eqnarray}
This equation, with the r.h.s. evaluated at some $\Sigma_{y}$, constitutes a working definition for the potential.

 We notice   that equations  (\ref{field equations for g}) and (\ref{D2 phi with residual term}) are identical to those of Brans-Dicke theory in $4D$ derived from the action 
\begin{equation}
\label{action in 4D}
S_{(4)} = \int{d^4 x \sqrt{ - g}\left[\phi R^{(4)} - \frac{\omega}{\phi} g^{\mu\nu}\left(D_{\mu}\phi\right)\left(D_{\nu}\phi\right) - V(\phi)\right]}  + 16 \pi \int{d^4 x \sqrt{- g} L_{m}^{(4)}}, 
\end{equation}
 with  $\sqrt{- g}\left[S_{\mu\nu} + T_{\mu\nu}^{(BD)}\right] \equiv 2 \delta S_{m}/\delta g^{\mu\nu}$, where $S_{m} = \int{d^4 x\sqrt{- g}L_{m}^{(4)}}$ represents the action for matter in $4D$.

The potential vanishes only in few particular cases. For example when $\omega = -1$, $y$ is a cyclic coordinate and $T_{A B}^{(5)} = 0$.  However,  if we are to recover a general version of BD theory in $4D$,  we must assume $\omega \neq -1$. Besides, the metric in $5D$ can depend on $y$. Therefore,  in general $V \neq 0$.

 \medskip

$\bullet$ Finally, for the line element (\ref{line element in 5D}) we have $R_{4\alpha}^{(5)} = \Phi P_{\alpha; \beta}^{\beta}$ \cite{WJPdeL} with 
\begin{equation}
P_{\alpha\beta} = \frac{1}{2 \Phi}\left(\stackrel{\ast}g_{\alpha\beta} - g_{\alpha\beta} g^{\mu\nu}\stackrel{\ast}g_{\mu\nu}\right).
\end{equation}
The dynamical equation for $P_{\alpha\beta}$ is obtained from (\ref{equation for the metric in 5D}) by  
setting $A = 4$, $B = \mu$, viz.,

\begin{equation}
\label{dynamical equation for P}
\left(\Phi \phi\right) P_{\mu; \alpha}^{\alpha} = 8 \pi T_{\mu 4}^{(5)} + \frac{\omega \stackrel{\ast}\phi D_{\mu}\phi}{\phi} + D_{\mu}\stackrel{\ast}\phi - \frac{1}{2}\stackrel{\ast}g_{\mu\lambda}D^{\lambda}\phi - \frac{\stackrel{\ast}\phi D_{\mu}\Phi}{\Phi}. 
\end{equation}
In the case where $T_{\mu 4}^{(5)} = 0$ and  $\phi = $ constant, this reduces to $P_{\mu; \alpha}^{\alpha} = 0$. In braneworld theory this quantity is proportional to the EMT of the matter on the brane \cite{equiv}. If $y$ is a cyclic coordinate, (\ref{dynamical equation for P}) reduces to $0 = 0$.

To keep contact with other works in the literature, we note that we can substitute  (\ref{equation for Phi})  in (\ref{definition of S}) to  obtain an expression where   the right hand side of (\ref{field equations for g}) only contains $T_{\mu\nu}^{(5)}$ as well as the derivatives of $\phi$, $\Phi$ and $g_{\mu\nu}$. In case that there is no $y$ dependence we neatly recover the formulae developed in \cite{Li}. However, our equations are completely different from those derived/used in \cite{Aguilar}.

\section{Scalar-vacuum Brans-Dicke cosmology in $5D$}

In cosmological applications the $5D$ metric (\ref{line element in 5D}) is commonly taken in the form
\begin{equation}
\label{cosmological metric in 5D, with y dependence}
dS^2 = n^2(t, y)dt^2 - a^2(t, y)\left[\frac{dr^2}{1 - k r^2} + r^2 \left(d\theta^2 + \sin^2 \theta d\varphi^2\right)\right] + \epsilon \Phi^2(t, y)dy^2,
\end{equation}
where $k = 0, + 1, - 1$ and $(t, r, \theta, \phi)$ are the usual coordinates for a spacetime with spherically symmetric spatial sections\footnote{We do not make any assumption about $\epsilon$. Rather  we let the FE to determine the signature of the extra dimension. In Cases $1$ and $5$ (with the exception of solution (\ref{General solution with alpha = 1, m = 1 and omega = omega1, s = - 3 l})) the extra dimension must be spacelike. All other solutions allow both signatures.}. 

For this line element the vacuum  $(T_{AB}^{(5)} = 0) $ Brans-Dicke field equations (\ref{equation for the metric in 5D}) reduce as follows:  
The temporal component $A = B = 0$ gives 
\begin{eqnarray}
\label{FE(00)}
3\frac{\dot{a}}{a}\left(\frac{{\dot{a}}}{a} + \frac{\dot{\Phi}}{ \Phi}\right) + \frac{3 k n^2}{a^2}  +\frac{3 \epsilon n^2}{\Phi^2}\left[\frac{\stackrel{\ast \ast}a}{a}  + \frac{\stackrel{\ast}a}{a}\left(\frac{\stackrel{\ast}a}{a} - \frac{\stackrel{\ast}\Phi}{\Phi}\right)\right]= \frac{1}{\phi}\left[\ddot{\phi} + \dot{\phi}\left(\frac{\omega \dot{\phi}}{2 \phi} - \frac{\dot{n}}{n}\right)\right] + \frac{\epsilon n^2 \stackrel{\ast}\phi}{\Phi^2 \phi}\left(\frac{\stackrel{\ast}n}{n} - \frac{\omega \stackrel{\ast}\phi}{2 \phi}\right);
\end{eqnarray}
the spatial components $A = B = 1, 2, 3$  reduce to 
\begin{eqnarray}
\label{FE(11)}
\frac{2\ddot{a}}{a} + \frac{\dot{a}}{a}\left(\frac{\dot{a}}{a} - \frac{2\dot{n}}{n}\right) + \frac{\ddot{\Phi}}{\Phi} + \frac{\dot{\Phi}}{\Phi}\left(\frac{2\dot{a}}{a} - \frac{\dot{n}}{n}\right) + \frac{k n^2}{a^2} +\frac{\epsilon n^2}{\Phi^2}\left[\frac{2\stackrel{\ast \ast }a}{a} + \frac{\stackrel{\ast}a}{a}\left(\frac{\stackrel{\ast}a}{a} + \frac{2 \stackrel{\ast}n}{n}\right) + \frac{\stackrel{\ast \ast}n}{n} - \frac{\stackrel{\ast}\Phi}{\Phi}\left(\frac{2\stackrel{\ast}a}{a} + \frac{\stackrel{\ast}n}{n}\right)\right]\nonumber \\
= \frac{\dot{\phi}}{ \phi}\left(\frac{\dot{a}}{a} - \frac{\omega \dot{\phi}}{2 \phi}\right) + \frac{\epsilon n^2 \stackrel{\ast}\phi}{\Phi^2 \phi}\left(\frac{\stackrel{\ast}a}{a} - \frac{\omega \stackrel{\ast}\phi}{2 \phi}\right);
\end{eqnarray}
the $A = B = 4$ component gives
\begin{eqnarray}
\label{FE(44)}
3\left[\frac{\ddot{a}}{a} + \frac{\dot{a}}{a}\left(\frac{\dot{a}}{a} - \frac{\dot{n}}{n}\right)\right] + \frac{3 k n^2}{a^2} + \frac{3 \epsilon n^2}{\Phi^2 }\left(\frac{\stackrel{\ast}a}{a}\right)\left(\frac{\stackrel{\ast}a}{a} + \frac{\stackrel{\ast}n}{n}\right) = 
\frac{\dot{\phi}}{\phi}\left(\frac{\dot{\Phi}}{\Phi} - \frac{\omega \dot{\phi}}{2 \phi}\right) + \frac{\epsilon n^2}{\Phi^2 \phi}\left[\stackrel{\ast \ast}\phi + \stackrel{\ast}\phi \left(\frac{\omega \stackrel{\ast}\phi}{2 \phi} - \frac{\stackrel{\ast}\Phi}{\Phi}\right)\right];
\end{eqnarray}
the mixed component $A = 0, B = 4$ yields
\begin{equation}
\label{FE(04)}
3\left(\frac{\stackrel{\ast}n \dot{a}}{n a} + \frac{\dot{\Phi}\stackrel{\ast}a}{\Phi a} - \frac{\stackrel{\ast}{\dot{a}}}{a}\right) = \frac{\stackrel{\ast}{\dot{\phi}}}{\phi} - \frac{\stackrel{\ast}n \dot{\phi}}{n \phi} - \frac{\dot{\Phi} \stackrel{\ast}\phi}{\Phi \phi} + \frac{\omega}{\phi^2}\dot{\phi}\stackrel{\ast}\phi.
\end{equation}
Finally, the wave equation (\ref{equation for phi}) becomes
\begin{equation}
\label{Dalambertian in 5D with y dependence}
\nabla^2 \phi = \frac{1}{n^2}\left[\ddot{\phi} + \dot{\phi}\left(\frac{3 \dot{a}}{a} + \frac{\dot{\Phi}}{\Phi} - \frac{\dot{n}}{n}\right) \right] + \frac{\epsilon}{\Phi^2}\left[\stackrel{\ast \ast}\phi + \stackrel{\ast}\phi\left(\frac{3 \stackrel{\ast}a}{a}  + \frac{\stackrel{\ast}n}{n} - \frac{\stackrel{\ast}\Phi}{\Phi}\right)\right] = 0.
\end{equation}

\subsection{Separation of variables}

In this section, we look for solutions to the above equations under the assumption that the metric coefficients are separable functions of their arguments.  In this framework, without loss of generality we can set\footnote{This is a consequence of  the freedom to perform the coordinate transformation $t = t(\bar{t})$, $y = y(\bar{y})$.} 
\begin{equation}
\label{separation of coordinates}
n(t, y) = N(y),\;\;\; a(t, y) = P(y)Q(t),\;\;\; \Phi(t, y) = F(t),\;\;\; \phi(t, y) = U(y)W(t). 
\end{equation}

Then, from (\ref{FE(04)}) we obtain
\begin{equation}
\label{FE(04) for separation of variables}
\frac{3 \dot{Q}}{Q}\left(\frac{\stackrel{\ast}N}{N} - \frac{\stackrel{\ast}P}{P}\right) + \frac{\dot{F}}{F}\left(\frac{3 \stackrel{\ast}P}{P} + \frac{\stackrel{\ast}U}{U}\right) + \frac{\dot{W}}{W}\left(\frac{\stackrel{\ast}N}{N} - \frac{\left(1 + \omega\right)\stackrel{\ast}U}{U}\right) = 0.
\end{equation}
Thus, separability  requires
\begin{equation}
\label{condition of compatibility}
\frac{\stackrel{\ast}N}{N} - \frac{\stackrel{\ast}P}{P} = c_{1} \frac{\stackrel{\ast}f}{f}, \;\;\;\;\frac{3 \stackrel{\ast}P}{P} + \frac{\stackrel{\ast}U}{U} = c_{2} \frac{\stackrel{\ast}f}{f}, \;\;\;\;\;\frac{\stackrel{\ast}N}{N} - \frac{\left(1 + \omega\right)\stackrel{\ast}U}{U}  = c_{3} \frac{\stackrel{\ast}f}{f},
\end{equation}
where $c_{1}$, $c_{2}$, $c_{3}$ are arbitrary constants and $f = f(y)$ is some function to be determined by the field equations. 

From (\ref{condition of compatibility}) we get

\begin{equation}
\label{NPU}
N = N_{0} f^{\left[\left(3 c_{1} + c_{2}\right)\left(\omega + 1\right) + c_{3}\right]/\left(3\omega + 4\right)}, \;\;\;P = P_{0} f^{\left[c_{2}\left(1 + \omega\right) - c_{1} + c_{3}\right]/\left(3\omega + 4\right)}, \;\;\;U = U_{0}f^{\left[c_{2} + 3\left( c_{1} - c_{3}\right)\right]/\left(3\omega + 4\right)},
\end{equation}
where $N_{0}$, $P_{0}$ and $U_{0}$ are constants of integration. Besides, from (\ref{FE(04) for separation of variables}) and (\ref{condition of compatibility}) it follows that 
\begin{equation}
\label{first condition on F, Q and W}
Q^{3 c_{1}} F^{c_{2}} W^{c_{3}} = \mbox{constant}.
\end{equation}

Now, to obtain an equation for $f$ we substitute (\ref{separation of coordinates}) and (\ref{NPU}) into the wave equation (\ref{Dalambertian in 5D with y dependence}). The requirement of separability yields a second-order differential equation for $f(y)$ whose first integral is 
\begin{equation}
\label{first integral for f}
\left(\frac{d f}{d y}\right)^ 2 = \left(C_{s} f^{2 c_{2}} + C\right) f^{2 \left[\left(3 - 3c_{1} - 4 c_{2} \right)\omega - 3c_{1} - 5 c_{2} - c_{3} +4\right]/\left(3\omega + 4\right)},
\end{equation} 
where $C_{s}$ is a separation constant  and $C$ is a constant of integration. This equation admits exact integration, in terms of elementary functions, in several cases. For example, setting $C = 0$ we get
\begin{equation}
\label{f}
f(y) \propto \left\{\begin{array}{cc}
            y^{\left(3\omega + 4\right)/\left[\left(3c_{1} + c_{2}\right)\left(\omega + 1\right) + c_{3}\right]},\;\;\;\left(3c_{1} + c_{2}\right)\left(\omega + 1\right) + c_{3} \neq 0, \\
e^{\pm \sqrt{C_{s}}y}, \;\;\;\;\left(3c_{1} + c_{2}\right)\left(\omega + 1\right) + c_{3} =  0, \;\;\;C_{s} > 0.
               \end{array}
\right.
\end{equation}
Similar power-law and exponential solutions can be obtained for $C_{s} = 0$ or $c_{2} = 0$.

Next, we substitute the metric functions (\ref{separation of coordinates}),  with (\ref{NPU}) and (\ref{first integral for f}), into the FE (\ref{FE(00)})-(\ref{FE(44)}). We find that these equations have the following structure

\begin{equation}
\label{reduced form}
H_{1}(t) + \eta_{0} C  H_{2}(t) \left[f(y)\right]^{- 2 c_{2}} + k \left(3\omega + 4\right) H_{3}(t) \left[f(y)\right]^{2 c_{1}} = 0,
\end{equation}
where $\eta_{0} \equiv  c_{2}^2 \left(5 + 4 \omega\right) + 2 c_{2}\left[c_{3} + 3c_{1}\left(1 + \omega\right)\right] - 3 \left(c_{3} - c_{1}\right)^2$;      $H_{1}(t)$ is a combination of $F(t)$, $Q(t)$ and $W(t)$ and their derivatives; $H_{2} \propto W^2 Q^2$; $H_{3} \propto W^2 F^2$. Inspection of the FE shows that the last two terms in (\ref{reduced form}) do not cancel out for $c_{2} = - c_{1} \neq 0$ and $F \propto Q$. Therefore, the assumption of separability demands either $c_{1} = 0$ for $k \neq 0$, or $k = 0$ for $c_{1} \neq 0$ as well as the  fulfillment of (at least) one of the following conditions: $C = 0$, $\eta_{0} = 0$, $c_{2} = 0$. 

A detailed analysis of the field equations is provided in the Appendix, where we derive the differential equations to be satisfied by the three functions $Q$, $F$ and $W$. Inspection of those equations reveals that they admit power-law and exponential solutions for several choices of the parameters. However, this is clear at the outset if we notice that   (\ref{FE(04) for separation of variables}) can also be written as

\begin{equation}
\frac{3 \stackrel{\ast}P}{P}\left(\frac{\dot{F}}{F} - \frac{\dot{Q}}{Q}\right) + \frac{\stackrel{\ast}N}{N}\left(\frac{3 \dot{Q}}{Q} + \frac{\dot{W}}{W}\right) +  \frac{ \stackrel{\ast}U}{U}\left(\frac{\dot{F}}{F} - \frac{\left(1 + \omega\right)\dot{W}}{W}\right) = 0,
\end{equation}
which requires
\begin{equation}
\label{QFW}
\frac{\dot{F}}{F} - \frac{\dot{Q}}{Q} = \bar{c}_{1}\frac{\dot{h}}{h},\;\;\;\;\; \frac{3 \dot{Q}}{Q} + \frac{\dot{W}}{W} = \bar{c}_{2}\frac{\dot{h}}{h} , \;\;\;\;\;\frac{\dot{F}}{F} - \frac{\left(1 + \omega\right)\dot{W}}{W} = \bar{c}_{3}\frac{\dot{h}}{h}, 
\end{equation}
where $\bar{c}_{1}$, $\bar{c}_{2}$, $\bar{c}_{3}$ are arbitrary constants and $h = h(t)$ is some function to be determined by the field equations.
Integrating (\ref{QFW}) we get
\begin{equation}
F \propto h^{\left[\left(3 \bar{c}_{1} +  \bar{c}_{2}\right)\left(\omega + 1\right) + \bar{c}_{3}\right]/\left(3\omega + 4\right)}, 
\;\;\;\;Q \propto h^{\left[\bar{c}_{2}\left(\omega + 1\right) - \bar{c}_{1} + \bar{c}_{3}\right]/\left(3\omega + 4\right)}, 
\;\;\;\;
 W \propto h^{\left[\bar{c}_{2} + 3\left(\bar{c}_{1} - \bar{c}_{3}\right)\right]/\left(3\omega + 4\right)}
\end{equation}
Following the same procedure as above,  from the wave equation (\ref{Dalambertian in 5D with y dependence}) we find
\begin{equation}
\label{first integral for h}
\left(\frac{d h}{d t}\right)^ 2 = \left(\bar{C}_{s} h^{2 \bar{c}_{2}} + \bar{C}\right) h^{2 \left[\left(3 - 3\bar{c}_{1} - 4 \bar{c}_{2} \right)\omega - 3\bar{c}_{1} - 5 \bar{c}_{2} - \bar{c}_{3} +4\right]/\left(3\omega + 4\right)},
\end{equation} 
where $\bar{C}_{s}$  and $\bar{C}$ are some new separation and integration constants, respectively. For several choices of the constants we obtain power-law and exponential solutions. For example, it we  set $\bar{C} = 0$ we get  

\begin{equation}
\label{h}
h(t) \propto \left\{\begin{array}{cc}
            t^{\left(3\omega + 4\right)/\left[\left(3\bar{c}_{1} + \bar{c}_{2}\right)\left(\omega + 1\right) + \bar{c}_{3}\right]},\;\;\;\left(3\bar{c}_{1} + \bar{c}_{2}\right)\left(\omega + 1\right) + \bar{c}_{3} \neq 0,\\
e^{\pm \sqrt{\bar{C}_{s}}t}, \;\;\;\;\left(3\bar{c}_{1} + \bar{c}_{2}\right)\left(\omega + 1\right) + \bar{c}_{3} =  0, \;\;\;\bar{C}_{s} > 0.
               \end{array}
\right.
\end{equation}
Once again, similar solutions are obtained ${\bar{C}}_{s} = 0$ or ${\bar{c}}_{2} = 0$.

In summary, the assumption of separation of variables generates four distinct classes  of solutions corresponding to the following choices:
\begin{eqnarray}
\label{Four families of solutions}
\mbox{I}&:& h(t) \propto t^{\left(3\omega + 4\right)/\left[\left(3\bar{c}_{1} + \bar{c}_{2}\right)\left(\omega + 1\right) + \bar{c}_{3}\right]}, \;\;\;\;f(y) \propto y^{\left(3\omega + 4\right)/\left[\left(3c_{1} + c_{2}\right)\left(\omega + 1\right) + c_{3}\right]}, \nonumber \\
\mbox{II}&:& h(t) \propto e^{\pm \sqrt{\bar{C}_{s}}t}, \;\;\;\;f(y) \propto y^{\left(3\omega + 4\right)/\left[\left(3c_{1} + c_{2}\right)\left(\omega + 1\right) + c_{3}\right]}, \nonumber \\
\mbox{III}&:& h(t) \propto t^{\left(3\omega + 4\right)/\left[\left(3\bar{c}_{1} + \bar{c}_{2}\right)\left(\omega + 1\right) + \bar{c}_{3}\right]}, \;\;\;\;f(y) \propto e^{\pm \sqrt{C_{s}}y},\nonumber \\
\mbox{IV}&:& h(t) \propto e^{\pm \sqrt{\bar{C}_{s}}t}, \;\;\;\;f(y) \propto e^{\pm \sqrt{C_{s}}y},
\end{eqnarray}
where the constants obey the conditions indicated in (\ref{f}) and (\ref{h}). When we substitute (\ref{Four families of solutions}) in the FE (\ref{FE(00)})-(\ref{Dalambertian in 5D with y dependence}) the latter reduce to a system of algebraic equations which provide the appropriate consistency relations  for the constants. 

\subsection{Power-law solutions}

 We now proceed to study in some detail the power-law solutions. We concentrate our attention on the spatially-flat scenario, which seem to be relevant to the present epoch of the universe \cite{BOOMERANG}.

We could use the analysis provided in  the Appendix and study case by case the differential equations for $Q$, $F$ and $W$. However, since we already know the form of the desired solutions it is much easier to start from  the family I in (\ref{Four families of solutions}).  For practical reasons it is convenient to  adopt another parameterization, namely  $\left(c_{1}, c_{2}, c_{3}\right) \rightarrow (\alpha, \beta, \gamma)$ and $\left(\bar{c}_{1}, \bar{c}_{2}, \bar{c}_{3}\right) \rightarrow (l, m , s)$, where  the power-law solution takes the form
\begin{equation}
\label{power law solution, general case}
n = A y^{\alpha}, \;\;\;a = B y^{\beta}t^{l}, \;\;\;\Phi = C t^{m}, \;\;\;\phi = D y^{\gamma}t^s. 
\end{equation}
Here $A, B, C, D$ are some constants with the appropriate units; while  $\alpha, \beta, \gamma$ and $l, m, s$ are parameters that  have to satisfy the field equations (\ref{FE(00)})-(\ref{Dalambertian in 5D with y dependence}).

Substituting (\ref{power law solution, general case}) into (\ref{Dalambertian in 5D with y dependence}) we obtain
\begin{equation}
\label{general compatibility condition}
s\left(s - 1 + 3 l + m\right) C^2 t^{(s - 2)}y^{(\gamma - 2\alpha)} + \epsilon \gamma \left(\gamma - 1 + 3\beta + \alpha\right) A^2 t^{(s - 2 m)}y^{(\gamma - 2)} = 0.
\end{equation}
The above is satisfied in five different cases
\begin{eqnarray}
\label{five different cases}
s &=& 0, \;\;\;\gamma = 0, \nonumber \\
s &=& 0, \;\;\; \gamma = 1 - 3\beta - \alpha,\nonumber \\
\gamma &=& 0,\;\;\; s = 1 - 3 l - m, \nonumber \\
s &=& 1 - 3 l - m, \;\;\; \gamma = 1 - 3 \beta - \alpha, \nonumber \\
 m &=& 1,\;\;\; \alpha = 1, \;\;\;s \left(s + 3 l\right)C^2 + \epsilon \gamma \left(\gamma + 3\beta\right)A^2 = 0.
\end{eqnarray}
We now proceed to study these cases with some detail. 

\subsubsection{Case $1$:} 

The selection   $s = 0$, $\gamma = 0$ corresponds to the usual general relativity in $5D$ with  $\phi =$ constant. In this case the power-law cosmological solutions are well-known in the literature \cite{well-known}. However, to make the paper self-consistent we provide them here. 

From (\ref{FE(04)}) we obtain $\alpha l + \beta \left(m - l\right) = 0$. Thus, assuming $l \neq m$ we get $\beta = \alpha l/(l - m)$. In this case the only power-law solution to the field equations is\footnote{We exclude here static universes where $l = \beta = 0$. These admit  both signatures $(\epsilon = \pm 1)$ and require one of the following: $\left(\alpha = 1, \; m = 1\right)$; $\left(\alpha = 0, \; m = 1\right)$; $\left(\alpha = 1, \; m = 0\right)$.} 
\begin{equation}
\label{old solution 1}
dS^2 = A^2 y^2 dt^2 - B^2 y^{2l/(l - 1)} t^{2 l}\left[d r^2 + r^2 \left(d\theta^2 + \sin^2\theta d\varphi^2\right)\right] - \frac{A^2 t^2}{\left(l - 1\right)^2}d y^2.
\end{equation}
(The field equations cannot be satisfied for $s = 0$, $\gamma = 0$ and $l = 1$, simultaneously). For $l = m$ the solution is 
\begin{equation}
\label{old solution 2}
dS^2 = \frac{A^2}{y} dt^2 - B^2 y \left[d r^2 + r^2 \left(d\theta^2 + \sin^2\theta d\varphi^2\right)\right] + \epsilon d y^2. 
\end{equation}
Here, contrary to what we see in (\ref{old solution 1}),  the coordinate $y$ is not restricted to be spacelike. Therefore,  by virtue of extra symmetry discussed in \cite{XtraSymmetry}, the transformation of coordinates $t = w^{3/2} \bar{y}$, $r = \bar{r}/\sqrt{w}$, $y =  w \bar{t}$ with $ w = \pm \sqrt{\epsilon}$ in (\ref{old solution 2}) generates the ``new" solution 
\begin{equation}
\label{old solution 3}
dS^2 = d{\bar{t}}^2 - B^2 \bar{t} \left[d {\bar{r}}^2 + {\bar{r}}^2 \left(d\theta^2 + \sin^2\theta d\varphi^2\right)\right] + \frac{\epsilon A^2}{\bar{t}} d {\bar{y}}^2. 
\end{equation}
We should note that although  the last two line elements   are diffeomorphic in $5D$ (allowing complex transformations of coordinates in $5D$), their interpretation in $4D$  is quite different\footnote{As discussed in section $2$, in this paper  we  follow the approach where our $4D$ spacetime is  recovered by going onto a hypersurface $\Sigma_{y}: y = y_{0} = $ constant, although other techniques are also possible \cite{reinventing}.}.

\subsubsection{Case $2$:} The choice $s = 0$ generates  solutions where the scalar field $\phi$ depends only on  $y$. Setting $s = 0$ and $\gamma = \left(1 - 3\beta - \alpha\right)$ in  (\ref{general compatibility condition}) we get $m =  3 l\left(\alpha - \beta\right)/\left(\alpha - 1\right)$, where $\alpha \neq 1$. Now, (\ref{FE(44)})  requires either $l = 1/2$ or $l = 0$. 

The solution for $l = 1/2$ is 
\begin{eqnarray}
\label{Case 2,  solution for l = 1/2}
dS^2 &=& A^2 y^{(1 + 3 \beta)/2}dt^2 - B^2 y^{2 \beta} t\left[d r^2 + r^2 \left(d\theta^2 + \sin^2\theta d\varphi^2\right)\right] + \frac{\epsilon C^2}{t} d y^2,\nonumber \\
\phi &=& D y^{3(1 - 5 \beta)/4}, \;\;\; \beta = \frac{14 + 15 \omega \pm 4 \sqrt{6\left(4 + 3 \omega\right)}}{94 + 75 \omega}.
\end{eqnarray}
 (When $\omega = - 94/75 \approx - 1.25$ the field equations require $\beta = 3/5$.)

The solution for $l = 0$ is
\begin{eqnarray}
\label{Case 2,  solution for l = 0}
dS^2 &=& A^2 y^{2 \alpha}dt^2 - B^2 y^{2 \beta} \left[d r^2 + r^2 \left(d\theta^2 + \sin^2\theta d\varphi^2\right)\right] + \epsilon d y^2,\nonumber \\
\phi &=& D y^{\left(1 - \alpha - 3\beta\right)}, \;\;\; \omega = - \frac{12 \beta^2 + 2\left(\alpha - 1\right)\left(3\beta + \alpha\right)}{\left(1 - \alpha - 3\beta\right)^2}.
\end{eqnarray}
It should be noted that this solution works perfectly well for $\alpha  = 1$ provided $\omega = - 4/3$. Also, when $\alpha = \left(1 - 3\beta\right)$ the denominator in $\omega$ vanishes  and $\phi =$ constant. What this suggests is that the choice  $\alpha = \left(1 - 3\beta\right)$ should reproduce  $5D$ general relativity. In fact, for this choice the field equations require either $\left(\alpha = - 1/2, \;\beta = 1/2\right)$ 
or $\left(\alpha = 1, \; \beta = 0\right)$, which yield  $\omega \rightarrow \infty$. For the former set of values  the Brans-Dicke solution (\ref{Case 2,  solution for l = 0}) reduces to (\ref{old solution 2}), while for the latter  it reduces to the third  static solution mentioned in footnote $5$.

\subsubsection{Case $3$:} The choice $\gamma = 0$ generates  solutions where the scalar field $\phi$ depends only on $t$. Setting $\gamma = 0$ and $s = \left(l - 3 l - m\right)$ in (\ref{general compatibility condition}) we get $\alpha = - 3\beta\left(l - m\right)/\left(m - 1\right)$, where $m \neq 1$. 
From (\ref{FE(00)}) it follows that either $\beta = 1/2$ or $\beta = 0$. The solution corresponding to $\beta = 1/2$ is
\begin{eqnarray}
\label{Case 3,  solution 1}
dS^2 &=& \frac{A^2}{y} dt^2 - B^2 y t^{2 l}\left[d r^2 + r^2 \left(d\theta^2 + \sin^2\theta d\varphi^2\right)\right] + \epsilon C^2 t^{(1 + 3 l)/2}d y^2,\nonumber \\
\phi &=& D t^{3(1 - 5 l)/4}, \;\;\; l = \frac{14 + 15 \omega \pm 4 \sqrt{6\left(4 + 3 \omega\right)}}{94 + 75 \omega}.
\end{eqnarray}
We note that this solution is related to (\ref{Case 2,  solution for l = 1/2}) by the transformation $t \leftrightarrow y$, $\beta \leftrightarrow l$, $\epsilon C^2 \leftrightarrow A^2$.

The solution for $\beta = 0$ is    
\begin{eqnarray}
\label{Case 3,  solution 2}
dS^2 &=& dt^2 - B^2 t^{2 l}\left[d r^2 + r^2 \left(d\theta^2 + \sin^2\theta d\varphi^2\right)\right] + \epsilon C^2 t^{2 m}d y^2,\nonumber \\
\phi &=& D t^{\left(1 - m - 3 l\right)}, \;\;\; \omega = - \frac{12 l^2 + 2\left(m - 1\right)\left(3 l + m\right)}{\left(1 - m - 3 l\right)^2}, 
\end{eqnarray}
which can formally be obtained from (\ref{Case 2,  solution for l = 0}) after the transformation 
$t = \left(w^{1 - \alpha}C/A\right)\bar{y}$, $r =  w^{- 2\beta} \bar{r}$, $y =   w \bar{t}$ with  $w = \pm \sqrt{\epsilon} $ and replacing  $\alpha$ and $\beta $ by $m$ and $l$, respectively. In the limit where $m \rightarrow \left(1 - 3 l\right)$, for which  $\phi =$ constant and $\omega \rightarrow \infty$, we recover the general-relativistic solutions (\ref{old solution 3}) and the second static metric mentioned in footnote $5$.

It should be noted that (\ref{Case 3,  solution 2}) is the general power-law solution for the case where the metric depends on time alone. There is a very particular case, given bellow by (\ref{solution with alpha = 0}) with $\beta = 0$, for which $\phi$ is allowed to depend on $y$.

\subsubsection{Case $4$}
We now look for solutions to the field equations of the form
 
\begin{eqnarray}
\label{general solution}
dS^2 &=& A^2 y^{2\left(1 - 3\beta - \gamma\right)} dt^2 - B^2 y^{2 \beta} t^{2 l}\left[d r^2 + r^2 \left(d\theta^2 + \sin^2\theta d\varphi^2\right)\right] + \epsilon C^2 t^{2\left(1 - 3 l - s\right)} d y^2,\nonumber \\
\phi &=& D y^{\gamma} t^{s}.
\end{eqnarray}
Substituting these expressions into (\ref{FE(00)}) we get
\begin{equation}
\label{omega for the general solution }
\omega = - \frac{2\left(\gamma^2 - \gamma + 3 \gamma \beta + 6 \beta^2 - 3\beta\right)}{\gamma^2}, \;\;\;\gamma \neq 0
\end{equation}
(If $\gamma = 0$ then we recover Case $3$). Now from (\ref{FE(04)}) we obtain
\begin{equation}
\label{s for the general solution}
s = \frac{\gamma^2\left(1 - 6 l\right) + 3\gamma \left[l + \beta\left(1 - 7l\right)\right]}{6\beta \left(1 - 2\beta\right) + \gamma\left(\gamma + 1\right)}.
\end{equation}
For these quantities we find that the choice $\beta = - \gamma/3$ $(s = - 3 l)$ produces a simple particular solution, which is given bellow by (\ref{General solution with alpha = 1, m = 1 and omega = omega1, s = - 3 l}).

The general solution to the field equations (\ref{FE(00)})-(\ref{Dalambertian in 5D with y dependence}), in the case under consideration, is generated by a quadratic equation for $l$. The explicit expression for $l$ is given by

\begin{eqnarray}
\label{l for the general solution}
l = \frac{9\left(6\beta + 7\gamma - 1\right)\beta^2 + 3\left(8\beta + \gamma - 2\right)\gamma^2 - 3\left(8\gamma + 3\right)\beta \pm \sqrt{3\left(3\beta + \gamma\right)\left(3 - 6\beta - \gamma\right)}\left[6\left(2\beta - 1\right)\beta - \left(\gamma + 1\right)\gamma \right]}{3\left(4 \gamma^3 - 13\gamma^2 + 7\gamma - 3\right)  + 9\left[ 11\left(6\beta + 5\gamma\right)\beta^2 - 15\left(3\beta - \gamma^2\right)\beta + 4\left(2 - 7\gamma\right)\beta\right] }.
\end{eqnarray}
Thus, (\ref{omega for the general solution })-(\ref{l for the general solution}) describe a two-parameter family of solutions. 
From a mathematical viewpoint, this expression might look cumbersome and/or non-interesting. However, from a physical viewpoint,  the fact that $l$ explicitly depends on $\beta $ and  $\gamma$ means  that the expansion rate of the universe, which is determined by $l$,  is manifestly determined by the  dynamics along the extra dimension.  

The above solution can be simplified if we add some physical requirements. The simplest ones arise from the choice of the coordinate/reference system.

\paragraph{Synchronous reference system:} The choice $g_{00} = 1$ is usual in cosmology; it corresponds to the so-called synchronous reference system where the coordinate $t$ is the proper time at each point. Thus, setting $A = 1$ and $\alpha = \left(1 - 3\beta - \gamma\right) = 0$ in 
(\ref{general solution}) the solution simplifies to
\begin{eqnarray}
\label{solution with alpha = 0}
dS^2 &=& dt^2 - B^2 y^{2 \beta} t^{2 l}\left[d r^2 + r^2 \left(d\theta^2 + \sin^2\theta d\varphi^2\right)\right] + \epsilon C^2 t^{2\left(1 - 3 l - s\right)} d y^2,\nonumber \\
\phi &=& D y^{\left(1 - 3 \beta\right)} t^{s},\;\;\;\omega = \frac{6 \beta\left(1 - 2\beta\right)}{\left(1 - 3\beta\right)^2},
\end{eqnarray}
where
\begin{eqnarray}
\label{coefficients for solution with alpha = 0}
l &=& \frac{3\left(3\beta^2 - 1\right) \pm \sqrt{3\left(2 - 3\beta\right)} \left[3 \beta\left(\beta + 1\right) - 2\right]}{3\left(3\beta^2 + 6\beta - 5\right)},\nonumber \\
s &=& \frac{\left(1 - 3\beta\right)\left[3 l\left(1 + \beta\right) - 1\right]}{3 \beta \left(1 + \beta\right) - 2}.
\end{eqnarray}
It is worthwhile to mention that there are no solutions having simultaneously $\alpha = 0$ and $\beta = 1/3$.
Besides, the condition  $\beta < 2/3$ ensures that $l$ is a real number and $\omega > - 4/3$. For completeness, we note that there is a unique  solution for which the metric depends only on time, while  the scalar field  is a function of both $t$ and $y$. Such a solution is generated by the choice $\beta = 0$ in the above expressions.

\paragraph{Gaussian normal coordinate system:} A popular choice in the literature is to use the five degrees of freedom to set $g_{4 \mu} = 0$ and $g_{44} = 1$. This is the  so-called Gaussian normal coordinate system based on $\Sigma_{y}$.    The solution in such coordinates can formally be obtained from (\ref{solution with alpha = 0})-(\ref{coefficients for solution with alpha = 0}) by changing $t \leftrightarrow y$, $l \leftrightarrow \beta$, $s \leftrightarrow \gamma$;  setting $\epsilon C^2 = A^2$;  and allowing the extra coordinate to have one or the other signature. The result is 
\begin{eqnarray}
\label{solution with m = 0}
dS^2 &=& A^2 y^{2\left(1 - 3\beta - \gamma\right)}dt^2 - B^2 y^{2 \beta} t^{2 l}\left[d r^2 + r^2 \left(d\theta^2 + \sin^2\theta d\varphi^2\right)\right] + \epsilon d y^2,\nonumber \\
\phi &=& D y^{\gamma} t^{\left(1 - 3 l\right)},\;\;\;\omega = \frac{6 l \left(1 - 2 l\right)}{\left(1 - 3 l \right)^2},
\end{eqnarray}
where
\begin{eqnarray}
\label{coefficients for solution with m = 0}
\beta &=& \frac{3\left(3 l^2 - 1\right) \pm \sqrt{3\left(2 - 3 l \right)} \left[3 l \left(l + 1\right) - 2\right]}{3\left(3 l^2 + 6 l - 5\right)},\nonumber \\
\gamma &=& \frac{\left(1 - 3 l\right)\left[3 \beta\left(1 + l\right) - 1\right]}{3 l \left(1 + l \right) - 2}.
\end{eqnarray}

$\bullet$ Let us now go back to the general solution (\ref{general solution}). 
For completeness we should consider the case where the denominator in (\ref{s for the general solution}) vanishes.  In such a case,  from (\ref{FE(04)}) and (\ref{s for the general solution}) we find that the parameters $\beta$ and $\gamma$ are restricted to be
\begin{eqnarray}
\label{allowed parameters when  the denominator in Case 4 vanishes }
\gamma &=& - 3, \;\;\; \beta = 1,\nonumber \\
 \gamma &=& \frac{6 l\left(5 l - 1\right)}{3 l^ 2 + 6 l - 1}, \;\;\;  \beta = \frac{l\left(9 l - 1\right)}{3 l^2 + 6 l - 1}, 
\end{eqnarray}
The first set of parameters  generates two families of solutions. Namely, 
\begin{eqnarray}
\label{solutions when the denominator in Case 4 vanishes}
dS^2 &=& A^2 y^2 dt^2 - B^2  y^{2}t^{2 l}\left[d r^2 + r^2 \left(d\theta^2 + \sin^2\theta d\varphi^2\right)\right] 
+ \epsilon C^2 t^{2 m_{(1,2)}}d y^2,\nonumber \\
\phi &=& D y^{- 3}t^{s_{(1,2)}}, \;\;\;\omega = - \frac{4}{3}, 
\end{eqnarray}
where $m_{(1)} = 1, s_{(1)} = - 3 l$ and $m_{(2)} = 3 l - 2, s_{(2)} = 3\left(1 - 2 l\right)$. For the second set of parameters in (\ref{allowed parameters when  the denominator in Case 4 vanishes }), the expression for $\omega$ given by (\ref{omega for the general solution }) simplifies 
to $\omega = \left(1 - 9 l\right)/2 l$ and the field equations have solutions only for two particular values of $l$. These are $l = 3/19$ and   $l \approx 0.153$, which is the real root of the cubic equation 
$\left(246 l^3 - 171 l^2 + 40 l  - 3\right) = 0$.

\subsubsection{Case $5$}
We now present the general solution for  $\alpha = 1$, $m = 1$. 
We assume $s \neq 0$, $\gamma \neq 0$, which were already considered in the previous cases. Thus, from  (\ref{five different cases}) 
we obtain\footnote{This case was recently discussed in \cite{Tae Hoon}. However, our solution (\ref{General solution with alpha = 1, m = 1 and omega = omega1}) seems to be much simpler.} 
\begin{equation}
\label{C for solution with alpha = 1, m = 1}
C^2 = - \frac{\epsilon \gamma \left(\gamma + 3\beta\right)}{s\left(s + 3l\right)}A^2, \;\;\;s \neq - 3 l.
\end{equation}
From (\ref{FE(04)}) we get
\begin{equation}
\label{beta for the general solution with alpha = 1, m = 1}
\beta = \frac{s\left[1 - \gamma\left(1 + \omega\right)\right] + 3 l + \gamma}{3\left(l - 1\right)}, \;\;\;l\neq 1.
\end{equation}
Now (\ref{FE(00)}) requires either
 \begin{equation}
\label{omega 1}
\omega = \omega_{1} = \frac{s^2\left(1 - \gamma\right) + s\left(3 l + \gamma\right) - 3 \gamma l\left(l - 1\right)}{\gamma s^2},  
\end{equation}
or
\begin{equation}
\label{omega 2}
\omega = \omega_{2} = \frac{4 s^2\left(1 - \gamma\right) + 2 s \left(1 + 3 l\right)\left(3 - \gamma\right) + 6 l\left(l + 1\right)\left(3 + \gamma\right)}{s \gamma\left(4 s + 3 + 9 l\right)}, \;\;\;4 s + 3 - 9 l \neq 0.
 \end{equation}

\paragraph{Solution with $\omega = \omega_{1}$:} Substituting (\ref{omega 1}) into (\ref{beta for the general solution with alpha = 1, m = 1}) we obtain $\beta = l \gamma /s$.  Consequently, from  (\ref{C for solution with alpha = 1, m = 1}) it follows that $C^2 = - \epsilon \left(\gamma^2/s^2\right) A^2$. Now, it can be verified that the field equations are satisfied by   
\begin{eqnarray}
\label{General solution with alpha = 1, m = 1 and omega = omega1}
dS^2 &=& A^2 y^2 dt^2 - B^2  y^{2\gamma l/s}t^{2 l}\left[d r^2 + r^2 \left(d\theta^2 + \sin^2\theta d\varphi^2\right)\right] -\frac{\gamma^2}{s^2} A^2 t^{2}d y^2,\nonumber \\
\phi &=& D y^{\gamma}t^{s}, \;\;\;\omega = \omega_{1}. 
\end{eqnarray}
The above solution works perfectly well for $l = 1$, although (\ref{beta for the general solution with alpha = 1, m = 1}) requires $l \neq 1$. Besides the extra coordinate has to be spacelike $(\epsilon = - 1)$. In addition, for the choice   $s = (l - 1)\gamma$ the line element (\ref{General solution with alpha = 1, m = 1 and omega = omega1}) becomes identical to the one in (\ref{old solution 1}), but now $\phi = D y^{s/(l - 1)} t^s$ and $\omega = (l - s)/s$. With this choice,  the $5D$ Brans-Dicke solution (\ref{General solution with alpha = 1, m = 1 and omega = omega1}) goes over the $5D$ general-relativistic solution (\ref{old solution 1}) in the limit\footnote{To avoid misunderstandings, we should mention that the limit $\omega \rightarrow \infty$ does not necessarily implies that we recover general relativity. As an  example consider the solution 
(\ref{Case 2,  solution for l = 1/2}): For $\omega \rightarrow \infty$ we get $\beta = 1/5$ and $\phi = $ constant, but the resulting metric does not satisfy the FE. A similar situation occurs in solutions 
 (\ref{solution with alpha = 0})-(\ref{coefficients for solution with alpha = 0}) and (\ref{solution with m = 0})-(\ref{coefficients for solution with m = 0}) for  $\beta \rightarrow  1/3$ and  $l \rightarrow 1/3$, respectively.}     $\omega \rightarrow \infty$ $(s \rightarrow 0)$.

For $s = - 3 l$  the solution is
\begin{eqnarray}
\label{General solution with alpha = 1, m = 1 and omega = omega1, s = - 3 l}
dS^2 &=& A^2 y^2 dt^2 - B^2  y^{- 2\gamma/3}t^{2 l}\left[d r^2 + r^2 \left(d\theta^2 + \sin^2\theta d\varphi^2\right)\right] 
+ \epsilon C^2 t^{2}d y^2,\nonumber \\
\phi &=& D y^{\gamma}t^{ - 3 l}, \;\;\;\omega = - \frac{4}{3}. 
\end{eqnarray}
We note that this is the only solution to the FE, with $\alpha = m = 1$,    which allows the extra coordinate to be timelike (see (\ref{old solution 1}), (\ref{General solution with alpha = 1, m = 1 and omega = omega1})).

\paragraph{Solution with $\omega = \omega_{2}$:} From the field equation (\ref{FE(44)}) we obtain the condition
\begin{equation}
\label{FE(44) requires for solution with alpha = 1, m = 1 and omega = omega2}
\left(s + 3 l\right)^2\left[\gamma \left(3 l + 1\right) + s \left(\gamma - 1\right)\right] = 0.
\end{equation}
Since $s + 3 l \neq 0$ (see (\ref{C for solution with alpha = 1, m = 1}) and (\ref{General solution with alpha = 1, m = 1 and omega = omega1, s = - 3 l})) it follows that
\begin{eqnarray}
s &=& - \frac{\gamma\left(3 l + 1\right)}{\gamma - 1}, \;\;\;\gamma \neq 1\nonumber \\
\gamma &=& 1, \;\;\;l = -\frac{1}{3}.
\end{eqnarray}
It can be verified that when $\omega = \omega_{2}$ the solution to the field equations is given by (\ref{General solution with alpha = 1, m = 1 and omega = omega1}) with the choice of parameters given above. 
When  the denominator in (\ref{omega 2}) vanishes, the solution is (\ref{General solution with alpha = 1, m = 1 and omega = omega1}) with  
\begin{equation}
\label{s when denominator in omega2 vanishes}
s = - \frac{3\left(1 + 3 l\right)}{4}.
\end{equation}
Thus, the choice $\omega = \omega_{2}$ does not generate new solutions to the FE. It just singles out some particular values for the parameters. Consequently,  (\ref{General solution with alpha = 1, m = 1 and omega = omega1}) is the general power-law solution with $\alpha = 1$, $m = 1$.

\section{Reduced Brans-Dicke cosmology in $4D$}

We now proceed to study the effective $4D$ picture generated by the $5D$ solutions discussed in the preceding section. 
For the line element (\ref{cosmological metric in 5D, with y dependence}),  the cosmological metric induced on spacetime hypersurfaces $\Sigma_{y}: y = y_{0} =$ constant  is just the spacetime part of (\ref{cosmological metric in 5D, with y dependence}), and the non-vanishing components of the induced Brans-Dicke energy-momentum tensor (\ref{T(BD)}) are
\begin{eqnarray}
\label{components of the Brans-Dicke theory}
\frac{8\pi}{\phi} {T_{0}^{0}}^{(BD)} &=& \frac{1}{n^2}\left(\frac{\ddot{\Phi}}{\Phi} - \frac{\dot{n}\dot{\Phi}}{n \Phi}\right) + \frac{\epsilon}{\Phi^2}\left[\frac{\stackrel{\ast \ast}n}{n} - 3\frac{\stackrel{\ast}a}{a}\left(\frac{\stackrel{\ast}a}{a} + \frac{\stackrel{\ast}\phi}{\phi}\right)  - \frac{\stackrel{\ast}n \stackrel{\ast}\Phi}{n \Phi}    + \frac{\omega {\stackrel{\ast}\phi}^2}{2\phi^2}\right] + \frac{V}{2 \phi},\nonumber \\
\frac{8 \pi}{\phi}{T_{1}^{1}}^{(BD)} &=& \frac{\dot{a}\dot {\Phi}}{n^2 a \Phi} + \frac{\epsilon}{\Phi^2}\left[\frac{\stackrel{\ast \ast}a}{a} - \frac{\stackrel{\ast}a}{a}\left(\frac{\stackrel{\ast}a}{a} + \frac{\stackrel{\ast}\Phi}{\Phi} +\frac{2 \stackrel{\ast}n}{n} + \frac{2 \stackrel{\ast}\phi}{\phi}\right) - \frac{\stackrel{\ast}\phi}{\phi}\left(\frac{\stackrel{\ast}n}{n} - \frac{\omega \stackrel{\ast}\phi}{2 \phi}\right)\right] + \frac{V}{2 \phi},
\end{eqnarray}
where $V = V(\phi)$ should be determined from (\ref{residual term}). We note that ${T_{2}^{2}}^{(BD)} = {T_{3}^{3}}^{(BD)} = {T_{1}^{1}}^{(BD)}$, which means that the induced EMT looks like a perfect fluid with energy density $\rho = {T_{0}^{0}}^{(BD)}$ and isotropic pressure $p = - {T_{1}^{1}}^{(BD)}$.

Let us first calculate the potential. With this aim we substitute the power-law line element (\ref{power law solution, general case}) into (\ref{residual term}) and evaluate on $\Sigma_{y}$.
We obtain
\begin{equation}
\label{evaluating V, general approach}
\left[\phi \frac{d V}{d\phi}\right]_{\mid_{\Sigma_{y}}} = - 2 m s \left(\omega + 1\right) D^{2/s}A^{-2}y_{0}^{2\left(\gamma - \alpha s\right)/s} \phi^{\left(s - 2\right)/s} - 
\epsilon {\cal{C}} D^{2 m/s} C^{- 2} y_{0}^{2\left(\gamma m - s\right)/s} \phi^{\left(s - 2 m\right)/s}, \;\;\;s \neq 0,
\end{equation}
where
\begin{equation}
\label{calC}
{\cal{C}} =  \gamma \omega\left[3\gamma + 2\left(\alpha + 3\beta - 1\right)\right] + 2\left[\gamma^2 - \gamma - 3\beta\left(\alpha + \beta\right)\right].
\end{equation}

$\bullet$ It turns out that ${\cal{C}}$ vanishes identically for all solutions discussed in Cases $3$ and $4$. Therefore in these cases, integrating (\ref{evaluating V, general approach}) and setting the constant of integration equal to zero we find

\begin{equation}
\label{potential for Cases 3 and 4}
V(\phi) = \left\{\begin{array}{cc}
            - \frac{2 m s^2 \left(\omega + 1\right) D^{2/s} y_{0}^{2\left(\gamma - \alpha s\right)/s}}{\left(s - 2\right) A^2} \phi^{\left(s - 2\right)/s}, \;\;\;s \neq 0, \;\;\;s \neq 2, \\
\\
          - 4 m \left(\omega + 1\right) D A^{- 2} y_{0}^{\left(\gamma - 2 \alpha\right)} \ln \phi,\;\;\;s = 2 .
               \end{array}
\right.
\end{equation}
Thus, the effective potential vanishes either in the particular case where $\omega = - 1$ or in the Gaussian normal coordinate system $(m = 0)$, i.e., for the solution (\ref{solution with m = 0})-(\ref{coefficients for solution with m = 0}).  

$\bullet$ In Case $5$, ${\cal{C}} \neq 0$ and from (\ref{evaluating V, general approach}) we get

\begin{equation}
\label{potential for Case 5}
V(\phi) = \left\{\begin{array}{cc}
           \frac{s \left[\omega s\left(s - 2 + 6 l\right) + 6 l\left(1 - 2 l\right)\right] D^{2/s} y_{0}^{2\left(\gamma - s\right)/s}}{\left(s - 2\right)A^{2}} \phi^{\left(s - 2\right)/s}, \;\;\;s \neq 0, \;\;\;s \neq 2,   \\
\\
          6 l \left(2\omega + 1 - 2 l\right) D A^{- 2} y_{0}^{\left(\gamma - 2 \alpha\right)} \ln \phi,\;\;\;s = 2.
               \end{array}
\right.
\end{equation}

$\bullet$ The parameter $s$ is equal to zero only in Cases $1$ and $2$. In Case $1$,  the general-relativistic $5D$ solution (\ref{old solution 1}) reduces on every hypersurface $\Sigma_{y}$ to the spatially flat FRW models of general relativity, viz., 

\begin{equation}
\label{spatially flat FRW models}
ds^2 = dS^2_{|\Sigma_{y}} = A^2 y_{0}^2 dt^2 - B^2 y_{0}^{2l/(l - 1)} t^{2 l}\left[d r^2 + r^2 \left(d\theta^2 + \sin^2\theta d\varphi^2\right)\right],
\end{equation}
 with 
\begin{equation}
\label{EMT for FRW models}
\rho = \rho_{0}\left(\frac{t_{0}}{t}\right)^2, \;\;\; p = n\rho,
\end{equation} 
where $\rho_{0} = 3 l^2/\left(8 \pi t_{0}^2 A^2 y_{0}^2\right)$ is the energy density measured at some time $t = t_{0}$ and 
\begin{equation}
\label{n for FRW models}
n = \frac{2 - 3 l}{3 l}.
\end{equation}

$\bullet$ In Case $2$,  to evaluate $V$ we substitute  (\ref{Case 2,  solution for l = 1/2}) into (\ref{residual term}). We find that $\phi d V/d \phi$ vanishes identically by virtue of the equation relating $\beta$ and $\omega$. The same is true for solution (\ref{Case 2,  solution for l = 0}). Therefore, without loss of generality we can set $V = 0$.  Now, from (\ref{components of the Brans-Dicke theory}) we find that  (\ref{Case 2,  solution for l = 1/2})  reduces on $\Sigma_{y}$ to a radiation dominated universe $(\rho = 3 p)$, regardless of $\omega$. Similarly, (\ref{Case 2,  solution for l = 0}) reduces  to empty space $(\rho = p = 0)$ on $\Sigma_{y}$.

\subsection{$4D$ cosmologies induced by the $5D$ solutions of Cases $3$ and $4$}

Cases $3$ and $4$, with $s \neq 2$,  give more general BD cosmological models\footnote{If $s = 2$ and $m \neq 0$, the logarithmic potential leads to a violation of the weak energy condition $\rho \geq 0$. Therefore, to avoid such a violation we should set $m = 0$ when $s = 2$, which in turns implies that $T_{\mu\nu}^{(BD)} = 0$ on every $\Sigma_{y}$.} in $4D$. In fact, from (\ref{components of the Brans-Dicke theory}), using (\ref{potential for Cases 3 and 4})  we get

\begin{eqnarray}
\label{TBD for cosmologies from  solutions 3 and 4}
\frac{8 \pi}{\phi}{T_{0}^{0}}^{(BD)} &=&  \frac{m\left[\left(m - 1\right)\left(s - 2\right) - s^2\left(\omega + 1\right)\right]}{\left(s - 2\right)y_{0}^{2\alpha} A^2 t^2},\nonumber \\
{T_{1}^{1}}^{(BD)} &=& -n {T_{0}^{0}}^{(BD)}, 
\end{eqnarray}
where 
\begin{equation}
\label{n for Cases 3 and 4}
n = \frac{s^2\left(\omega + 1\right) - l\left(s - 2\right)}{\left(m - 1\right)\left(s - 2\right) - s^2\left(\omega + 1\right)}.
\end{equation}
Thus, the effective pressure and density satisfy the barotropic equation of state
\begin{equation}
p = n \rho.
\end{equation}
 We immediately notice that the $5D$ solutions (\ref{solution with m = 0})-(\ref{coefficients for solution with m = 0}) in the Gaussian normal frame $(m = 0)$ reduce to scalar-vacuum cosmologies in $4D$.
One can verify that for all solutions in Cases $3$ and $4$,  $n$ can be written as
\begin{equation}
\label{general form of n}
n = \frac{2 - 3 l - s}{3 l},
\end{equation}
and
\begin{equation}
\label{m for Cases 3 and 4}
m = 1 - 3 l - s = - 1 + 3 l n.
\end{equation}
We can use these expressions to eliminate $s$ and $m$ in (\ref{TBD for cosmologies from  solutions 3 and 4})-(\ref{n for Cases 3 and 4}), viz.,  

\begin{eqnarray}
\label{omega in terms of l and n for Cases 3 and 4}
\omega &=& \frac{6 l\left(2 + 3 n\right) - 6 l^2\left[3 n \left(n + 1\right) + 2\right] - 4}{\left[3 l\left(n + 1\right) - 2\right]^2}, \nonumber \\
\frac{8 \pi}{\phi}{T_{0}^{0}}^{(BD)} &=& \frac{\left(l + 2 - 3 l n\right)\left(1 - 3 l n\right)}{\left(n + 1\right)A^2 y_{0}^{2 \alpha} t^2}, \;\;\; {T_{1}^{1}}^{(BD)} = - n {T_{0}^{0}}^{(BD)}.
\end{eqnarray}
As a consequence, we can  formulate the reduced Brans-Dicke  cosmological models in $4D$, derived  from the $5D$ solutions of Cases $3$ and $4$,   solely in terms of $l$ and $n$ as

\begin{eqnarray}
\label{cosmologies in 4D from Cases 3 and 4}
ds^2 &\equiv& dS^2_{|\Sigma_{y}} = A^2 y_{0}^{2 \alpha}dt^2 - B^2 y_{0}^{2 \beta} t^{2 l}\left[dr^2+ r^2 \left(d\theta^2 + \sin \theta d\varphi^2\right)\right],\nonumber \\
\rho &=&  \rho_{0} \left(\frac{t_{0}}{t}\right)^{3 l (n + 1)}, \nonumber \\
\phi &=& \phi_{0} \left(\frac{t}{t_{0}}\right)^{[2 - 3 l\left(n + 1\right)]},\nonumber \\
V &=& V_{0}\left(\frac{\phi}{\phi_{0}}\right)^{3 l\left(n + 1\right)/\left[3 l\left(n + 1\right) - 2\right]},
 \end{eqnarray}
with 
\begin{eqnarray}
\label{The constants for the 4D cosmologies generated by  Cases 3 and 4}
\phi_{0} &=& \frac{8 \pi \left(n + 1\right) \rho_{0} {\bar{t}}_{0}^2 }{\left(l + 2 - 3 l n\right)\left(1 - 3 l n\right)}, \;\;\;{\bar{t}}_{0} \equiv A y_{0}^{\alpha} t_{0} \nonumber \\
V_{0} &=& \frac{16 \pi \rho_{0}\left[l \left(1 + 3 n^2\right) - 2 n\right]}{ 2 + l\left(1 - 3 n\right) },   
\end{eqnarray}
where $\rho_{0}$ refers to the value of the energy density at some arbitrary fixed time $t_{0}$,  
and 
\begin{equation}
\label{l for solutions 3 and 4}
l = l_{(\pm)} = \frac{6\left(\omega + 1\right) + 3 n\left(2\omega + 3\right) \pm \sqrt{9 n^2 + 12 \left(1 + \omega\right)\left(3 n - 1\right)}}{9\left(\omega + 2\right)n^2 + 18\left(\omega + 1\right) n + 3\left(3 \omega + 4\right)}.
\end{equation} 
We recall that, in principle, $l$ is an observable quantity because it is related to the acceleration parameter $q = - a \ddot{a}/\dot{a}^2$ by 
\[l = \frac{1}{q + 1}.\]
 
The denominator of $\omega $ in (\ref{omega in terms of l and n for Cases 3 and 4}) vanishes when $l = 2/3\left(n + 1\right)$ and $\omega \to \mbox{sign}\left(3 n - 1\right) \times \infty$. In this limit $s = 0$, $\phi = $ constant and  from (\ref{m for Cases 3 and 4}) it follows that $m = \left(n - 1\right)/\left(n + 1\right)$, which means that the extra dimension contracts, or compactifies $(m < 0)$, for any $- 1 < n < 1$. Consequently,  formally for $|\omega| \rightarrow \infty$ we recover the  usual spatially flat FRW cosmology of ordinary general relativity.

\paragraph{Matter-dominated universe:}Since the present epoch of the universe is matter dominated let us consider the case where $n = 0$. In this case the weak energy condition is satisfied for any $l > 0$. On the other hand, the condition $\omega > - 3/2$ restricts the range of $l$ to be  either $l < 2\left(1 - \sqrt{2/3}\right) \approx 0.37$ 
or $l > 2\left(1 + \sqrt{2/3}\right) \approx 3.633$. The former range requires $q > \left(1 + \sqrt{6}\right)/2 \approx 1.72$, which is inapplicable to the present epoch. However, the latter range leads to 
 $q <  \left(1 - \sqrt{6}\right)/2 \approx - 0.72$, which is within the current observed measurements $q = - 0.67 \pm 0.25$ \cite{Wendy}.  We note that  the parameter $l$ is a real number only if $\left(1 + \omega\right) \leq 0$. 

\paragraph{Radiation-dominated universe:} For $n = 1/3$ we find that ${T_{0}^{0}}^{(BD)} > 0$ in (\ref{omega in terms of l and n for Cases 3 and 4}) requires $l < 1$, i.e., $q > 0$. Now from (\ref{l for solutions 3 and 4}) we get $l_{(+)} = 1/2$ and $l_{(-)} = 2\left(1 + \omega\right)/\left(4\omega + 5\right)$. The solution $l_{(+)} = 1/2$ is identical to the radiation-dominated epoch of general relativity, 
regardless of the specific value of $\omega$. The second solution violates the condition $0 < l < 1$ for any $\omega < - 1$, as required by the dust model. 

\bigskip

Therefore, the effective BD cosmology in $4D$  (\ref{cosmologies in 4D from Cases 3 and 4})-(\ref{l for solutions 3 and 4}) gives a decelerated radiation era, which is consistent with the big-bang nucleosynthesis scenario of general relativity $(l = 1/2)$,  as well as an accelerating matter dominated era compatible with present observations.

\bigskip 

To avoid misunderstanding, we should mention that the above discussion assumes that the parameters $n$ and $\omega$ are independent, which is true in the $4D$ cosmologies generated by (\ref{Case 3,  solution 2}) and (\ref{general solution}): giving $n$ and $\omega$ we obtain $l$ from (\ref{l for solutions 3 and 4}). This in turn defines the rest of the 
solution (\ref{cosmologies in 4D from Cases 3 and 4})-(\ref{The constants for the 4D cosmologies generated by  Cases 3 and 4}).
The $5D$ solutions (\ref{Case 3,  solution 1}) and (\ref{solution with alpha = 0}) only have one independent parameter. Therefore they lead to a more restricted class of BD cosmologies in $4D$. In particular, one can show that (\ref{Case 3,  solution 1}) yields shrinking (instead of expanding) dust models. Regarding the solutions in the synchronous frame (\ref{solution with alpha = 0}), they produce $4D$ models of expanding dust, but do not yield  accelerated expansion.

\medskip

$\bullet$ If we assume $V = 0$ at the beginning, then we obtain a set  of BD cosmologies in $4D$ which precludes the existence of a dust-filled universe. In fact, in  
the case where $V = 0$, $(m \neq 0,  \omega = - 1)$
 instead of (\ref{TBD for cosmologies from  solutions 3 and 4}) we have 
\begin{eqnarray}
\label{V = 0, Cases 3 and 4}
\frac{8 \pi}{\phi}{T_{0}^{0}}^{(BD)} &=&  \frac{m\left(m - 1\right)}{t^2 \left(A y_{0}^{\alpha}\right)^2},\nonumber \\
\frac{8 \pi}{\phi}{T_{1}^{1}}^{(BD)} &=& \frac{l m}{t^2 \left(A y_{0}^{\alpha}\right)^2}. 
\end{eqnarray}
Thus, a dynamical  universe $(l \neq 0)$ with $p = 0$ requires $\rho = 0$.

\subsection{$4D$ cosmologies induced by the $5D$ solutions of Case $5$}

Finally, let us study the effective $4D$ world generated by the $5D$ solutions discussed in Case $5$. We emphasize that these are the only solutions demanding the extra dimension to be spacelike. 

Now the effective potential is given by (\ref{potential for Case 5}). To avoid contradictions with the weak energy condition we disregard the logarithmic potential (see footnote $9$). In this context, from (\ref{components of the Brans-Dicke theory}) we find the effective EMT  as
\begin{eqnarray}
\label{EMT for the solutions of Case 5}
\frac{8\pi}{\phi} {T_{0}^{0}}^{(BD)}  = \frac{3 l\left[s\left(s + 3 l\right))+ \gamma l\left(1 - s - 3 l\right)\right]}{y_{0}^2 t^2  \gamma \left(s - 2\right)A^2},\;\;\;{T_{1}^{1}}^{(BD)} =  - n  {T_{0}^{0}}^{(BD)}, 
\end{eqnarray}
where $n$ is the same as in Cases $3$ and $4$, namely
\[
n = \frac{2 - 3 l - s}{3 l}.\]
As a consequence, from (\ref{omega 1}) we find that $\gamma$ can be expressed as
\begin{eqnarray}
\label{gamma and s for Case 5}
\gamma &=& \frac{\left(3 l n - 2\right)\left[3 l\left(n + 1\right) - 2\right]}{3 l\left(n + 1\right)\left[3 l \left(n + 1\right)\left(1 + \omega\right) - 4\omega - 3\right] + 3 l \left(l - 1\right) + 2\left(1 + 2\omega\right)}.
\end{eqnarray}
Substituting this into (\ref{EMT for the solutions of Case 5}) we get
 
\begin{eqnarray}
\frac{8\pi}{\phi} {T_{0}^{0}}^{(BD)}  &=& \frac{2\left(2 l - 1 - 2\omega\right) + 3 l\left(n + 1\right)\left[4\omega + 3 - l - 3 l\left(n + 1\right)\left(\omega + 1\right)\right]}{y_{0}^2 t^2 (n + 1) A^2}.
\end{eqnarray}
After some calculations we find that the effective cosmology in $4D$ is formally identical  to (\ref{cosmologies in 4D from Cases 3 and 4}) except for the fact that now\footnote{Here $\phi_{0}$ and $V_{0}$ depend on three independent parameters, $n$, $l$ and $\omega$. If one fixes  $\omega$ as in  (\ref{omega in terms of l and n for Cases 3 and 4}), then (\ref{phi0 and V0 for Case 5}) reduces to  (\ref{The constants for the 4D cosmologies generated by  Cases 3 and 4}).}  
\begin{eqnarray}
\label{phi0 and V0 for Case 5}
\phi_{0} &=& \frac{8 \pi \left(n + 1\right)\rho_{0} {\bar{t}}_{0}^2 }{2\left(2 l - 1 - 2\omega\right) + 3 l\left(n + 1\right)\left[4\omega + 3 - l - 3 l\left(n + 1\right)\left(\omega + 1\right)\right]}, \;\;\;{\bar{t}}_{0} \equiv A y_{0}t_{0}\nonumber 
\\ \nonumber \\
V_{0} &=& \frac{8\pi \rho_{0}\left[3 l\left(n + 1\right) - 2\right]\left\{2\left[1 + \omega \left(1 - n\right)\right] + l\left[3\omega\left(n^2 - 1\right) - 4\right]\right\}}{2\left(2 l - 1 - 2\omega\right) + 3 l\left(n + 1\right)\left[4\omega + 3 - l - 3 l\left(n + 1\right)\left(\omega + 1\right)\right]}.
\end{eqnarray}
Let us consider the above solution for large values of $\omega$, in accordance  with the solar system bound on $\omega (> 600)$. From the above expressions it follows that $\rho > 0$, $\phi > 0$ require $l$ in the range
\[
l = \frac{2}{3\left(n + 1\right)} \pm \frac{2}{3\sqrt{3}\left(n + 1\right)^{3/2}\sqrt{\omega}} + O(\frac{1}{\omega}).  
\]
Consequently, as $\omega \to \infty$ we recover the usual spatially flat FRW cosmologies of general relativity.

Let us now consider  a matter-dominated universe $(n = 0)$. Using that $l = 1/(q + 1)$ we find that the positivity of the effective density requires 
\[
\omega <  \frac{9 q - 2 q^2 - 1}{\left(2 q - 1\right)^2}.
\]
For the sake of argument, let us consider an accelerated expansion with $q = - 2/3$, which corresponds to $l = 3$. This requires $\omega < - 71/49 \approx - 1.449$.  In general,  $q < 0$ requires $\omega < - 1$. As $q \to 1/2$, which is the general-relativistic value of $q$ for dust, the theory becomes consistent with increasingly larger values of $\omega$.

We note that the solution can give accelerating matter-dominated era as well as decelerating radiation-dominated era  $(n = 1/3)$ for the same value of $\omega$ in the range $- 3/2 \leq \omega < - 1$. For example if we choose $\omega = - 1.45$, then a consistent model of  decelerating radiation era is obtained for   $l < 0.88$,  and accelerated dust era for $l > 3$, approximately.

\medskip

$\bullet$ For completeness, let us consider the case where $V = 0$. Coming back to the original expression for $\phi dV/d\phi$, we find that $V = 0$ requires  
\begin{equation}
\gamma = \frac{s\left(s + 6 l - 2\right)\left(s + 3 l\right)}{\left(s + 3 l - 1\right)\left[s^2 + s\left(3 l - 2\right) + 6 l\left(l - 1\right)\right]}.
\end{equation} 
Substituting into (\ref{omega 1}) we obtain
\begin{equation}
\omega = \frac{6\left(2 l - 1\right) l}{s\left(s + 6 l - 2\right)}.
\end{equation}
Thus, for $V = 0$ (omitting intermediate calculations) we find

\begin{equation}
\label{V = 0, Case 5}
p = n \rho, \;\;\;n = \frac{2 - s - 3l}{3 l}, \;\;\;\omega = \frac{2\left(2 l - 1\right)}{\left(1 - n\right)\left[2 - 3 l\left(n + 1\right)\right]}, \;\;\;\frac{8\pi}{\phi} {T_{0}^{0}}^{(BD)} = \frac{2 - l + 3 n l \left[l \left(3 n + 1\right) - 3\right]}{y^2 A^2 t^2 \left(1 - n\right)}.
\end{equation}
Consequently, the effective  BD cosmology  in $4D$ can be written as 
\begin{eqnarray}
\label{original BD dust solution}
\phi &=& \phi_{0}\left(\frac{t}{t_{0}}\right)^{2\left(1 - 3 n\right)/\left[4 + 3\omega\left(1 - n^2\right)\right]},\nonumber \\
a &=& a_{0} \left(\frac{t}{t_{0}}\right)^{2\left[1 + \omega\left(1 - n\right)\right]/\left[4 + 3\omega\left(1 - n^2\right)\right]}, 
\end{eqnarray}
with
\begin{equation}
\label{Generalizing BD solutions}
\phi_{0} = \frac{4\pi \rho_{0} {{\bar{t}}_{0}}^2 \left[4 + 3\omega \left(1 - n^2\right)\right]^2 }{\left(3 + 2 \omega\right)\left[4 - 6 n + 3\omega \left(n - 1\right)^2\right]}, \;\;\;{\bar{t}}_{0} \equiv A y_{0}t_{0}.
\end{equation}
In this parameterization, the  $n = 0$ models become identical to those  presented in the original BD paper \cite{Dicke}.  For $n \neq 0$ we recover the type A-I solutions discussed in \cite{Chakraborty}, although in a slightly different notation. From (\ref{V = 0, Case 5}) 
it is easy to see that any $\omega$ in  the range $- 2 < \omega <  -3/2$ $(1 < l < 2)$ yields a matter-dominated universe with accelerated expansion. However, this range is inconsistent with a radiation dominated epoch, which demands $\omega = - 3/2$ $(l = 1/2)$.  

\section{Summary}

Brans-Dicke theory in $4D$ can explain the observed accelerated expansion of the present matter-dominated universe, without invoking the presence of dark energy,  if $\omega$ is allowed to take  some value in the range $- 2 < \omega <  -3/2$. However,  $\omega$ in this range does not produce a consistent radiation-dominated  epoch  with decelerating expansion, as requires the big-bang nucleosynthesis scenario. One way out of this problem is to introduce a self interacting potential \cite{Bertolami}. Another way is to consider a modified BD theory with a varying  $\omega$  \cite{Pavon}. Both approaches are not free of criticism. In one of them the potential is added by hand, while the other creates  the necessity  of finding  the fundamental  mechanism driving the variation of $\omega$.  

A third way out of this problem is to resort  to higher dimensions. In \cite{Li} it was shown that the BD field equations in $5D$ can be reduced,  on a  hypersurface orthogonal to the extra dimension,    to those of  GR in $4D$  coupled to two scalar fields. These fields may account for the late-time accelerated expansion provided several other conditions are met: the extra coordinate is cyclic; $T_{A B}^{(5)} \neq 0$; $T_{A B}^{(5)}\hat{n}^{A} =  0$;  $a^3 \Phi^m = $ constant, where $m$ is a positive constant;  the effective matter in $4D$ has negligible pressure.    

In this work we have presented an approach  which is free of the above criticism and conditions.  Without making any assumption on the nature of the extra coordinate or the matter content in $5D$, we have shown that the BD field equations in $5D$ are equivalent to those of BD in $4D$ derived from the action (\ref{action in 4D}) with non-vanishing scalar potential $V = V(\phi)$. The potential is not introduced by hand, instead the reduction procedure provides an expression, namely (\ref{residual term}),  that determines the shape of $V(\phi)$ up to a constant of integration [(\ref{potential for Cases 3 and 4})-(\ref{potential for Case 5})]. It also establishes   the explicit formulae for the effective EMT in $4D$ [(\ref{definition of S})-(\ref{T(IMT)})]. This extends and generalize some previous results recently obtained by the present author \cite{LateAcceleration}.

In the context of FRW cosmological models we have integrated the vacuum FE in $5D$ under the sole assumption of separation of variables. We analyzed in detail the class of power-law solutions (\ref{power law solution, general case}). We obtained  a large family of solutions,  namely (\ref{General solution with alpha = 1, m = 1 and omega = omega1}),  that has three free parameters;  various families with  two free parameters [(\ref{Case 2,  solution for l = 0}), (\ref{Case 3,  solution 2}), (\ref{general solution})]; one free parameter [(\ref{Case 2,  solution for l = 1/2}), (\ref{Case 3,  solution 1}), (\ref{solution with alpha = 0}), (\ref{solution with m = 0})], and  solutions that only exist for $\omega = - 4/3$ [(\ref{solutions when the denominator in Case 4 vanishes}), (\ref{General solution with alpha = 1, m = 1 and omega = omega1, s = - 3 l})].  Thus, the spectrum of BD power-law solutions in $5D$ is significantly larger than the one  of GR in $5D$, which can have at most one free parameter (Case $1$).

We discussed the effective $4D$ world generated by our solutions. We found that the theory yields power-law and logarithmic potentials, except for the solutions derived in  Case $2$ and those in the Gaussian normal  frame where we can set $V = 0$, without loss of generality.  Certainly,   one can assume $V = 0$ at the outset in (\ref{residual term}), in which case the theory in $4D$ would be identical to the original BD \cite{Dicke}. However, such assumption imposes a strong  constraint on the parameters of the solutions leading to restricted cosmological models in $4D$ which have no, or little, relevance to the problem of cosmic acceleration [(\ref{V = 0, Cases 3 and 4}), (\ref{V = 0, Case 5})].

We found that all BD models in $4D$ can formally be expressed as in (\ref{cosmologies in 4D from Cases 3 and 4})  with $\phi_{0}$ and $V_{0}$ given by (\ref{The constants for the 4D cosmologies generated by  Cases 3 and 4}) or (\ref{phi0 and V0 for Case 5}). The models are given in terms of the parameters $n$, $l$ and $\omega$. When at least two of them are independent, which occurs for a large number of solutions [(\ref{Case 2,  solution for l = 0}), (\ref{Case 3,  solution 2}), (\ref{general solution}), (\ref{General solution with alpha = 1, m = 1 and omega = omega1})] we can obtain models that for the same $\omega$ can give the present accelerated expansion and a decelerated radiation-dominated epoch as required by primordial nucleosynthesis.  

We should mention   that as in the conventional $4D$ Brans-Dicke theory,  in our models the (effective) EMT obeys the ordinary  conservation law $D_{\mu} {T^{\mu}_{\nu}}^{(BD)} = 0$ (the same as in Einstein's theory), which in the cosmological realm yields  the usual equation of motion 
\begin{equation}
\label{conservation equation}
\dot{\rho} + 3\frac{\dot{a}}{a}\left(\rho + p\right) = 0.
\end{equation}

Although, we do not require the introduction of a self interacting potential in $4D$ or matter in $5D$, the range of $\omega$ obtained in our work is consistent with the one  obtained by a number of authors in the literature  in the context of  several cosmological models and  different versions of BD theory in $4D$ and $5D$ \cite{Bertolami}-\cite{Li 1}: In all cases cosmic accelerated expansion requires $\omega < - 1$.

There is also evidence in the literature that small values of $|\omega|$ are required by inflationary models \cite{La} and structure formation \cite{Gaztanaga} in scalar-tensor theory. Certainly, these theoretical results contradict solar system experiments which impose the constraint $\omega > 600$. However, this solar system limit does not necessarily imply that the evolution of the universe is, at all scales, close to general relativity. The fact is that GR is poorly tested on cosmic scales \cite{Rachel} and no experiment has been done to test BD in cosmological scale yet.  Consequently,  one cannot discard scalar-tensor scenarios of the sort discussed here and in \cite{Bertolami}-\cite{Li 1}, \cite{La}-\cite{Gaztanaga}   on the basis that in cosmological scales $\omega$  does not meet solar system requirements. 

Thus, the problem of (in)compatibility  between astronomical and cosmological requirements remains open. Our recent work \cite{LateAcceleration} indicates that a variable (homogeneous) $\omega$ cannot solve the problem. Perhaps, local inhomogeneities might give rise to high values of $\omega$ consistent with astronomical observations \cite{Pavon}.

\renewcommand{\theequation}{A-\arabic{equation}}
  \setcounter{equation}{0}  
  \section*{Appendix A: Integrating the field equations}  

In section $3$ we have seen that the scalar-vacuum Brans-Dicke FE for the cosmological metric (\ref{cosmological metric in 5D, with y dependence}) 
give five partial differential equations to be solved for four unknowns, $n(t, y)$, $a(t, y)$, $\Phi(t, y)$ and $\phi(t, y)$. The assumption that the metric is separable increases the number of unknowns to six. The aim of this appendix is to show, by means of explicit integration,  the consistency of this assumption with the 
field equations.

\paragraph{I:}
First, we assume that $c_{1} \neq  0$.  Then from (\ref{first condition on F, Q and W}) we get 

\begin{equation}
\label{equation for Q}
Q = Q_{0} F^{- \left(c_{2}/3 c_{1}\right)} W^{- \left(c_{3}/3 c_{1}\right)}, 
\end{equation}
where $Q_{0}$ is a constant. Substituting this and (\ref{first integral for f}) into the wave equation (\ref{Dalambertian in 5D with y dependence}) we obtain 
\begin{equation}
\label{equation for ddot W, c1 neq 0}
c_{1}F W \ddot{W} - \dot{W} \left[c_{3} F \dot{W} +  \left(c_{2} - c_{1}\right) W \dot{F}\right] + \frac{\epsilon c_{1} c_{2} C_{s}N_{0}^2\left[c_{2} + 3 \left(c_{1} - c_{3}\right)\right]}{\left(3\omega + 4\right)} F^{- 1} W^{2} = 0.
\end{equation}
When $c_{1} \neq 0$ the assumption of separability requires $k = 0$ and the fulfillment of one of the following conditions (i) $C = 0$, (ii) $\eta_{0} = 0$, (iii) $c_{2} = 0$ (cases Ia and Ib bellow).

 \paragraph{Ia:} Now we substitute the metric functions (\ref{separation of coordinates}), including (\ref{NPU}), (\ref{first integral for f}) and (\ref{equation for Q}), into the FE (\ref{FE(00)})-(\ref{FE(44)}) with $k = 0$ and either $C = 0$ or $\eta_{0} = 0$. We find that (\ref{FE(00)}) is a second-order differential equation for   $W$, while (\ref{FE(11)}) and (\ref{FE(44)}) contain the second derivatives of $W$ and $F$.  Next, we  isolate $\ddot{W}$ from (\ref{equation for ddot W, c1 neq 0}) and substitute it into (\ref{FE(00)}) to obtain the following first-order differential equation
\begin{equation}
\label{first-order equation for W and F}
\frac{{\hat{c}}_{1} \dot{W}^2}{W^2} + \frac{ {\hat{c}}_{2} \dot{W} \dot{F}}{W F} + \frac{{\hat{c}}_{3}\dot{F}^2}{F^2} + \frac{{\hat{c}}_{4}}{F^2} = 0
\end{equation}
with 
\begin{eqnarray}
{\hat{c}}_{1} &\equiv& 3 c_{1}^2 \omega + 6 c_{1}c_{3} - 2c_{3}^2, \nonumber \\
{\hat{c}}_{2} &\equiv& 6 c_{1}\left(c_{2} + c_{3} - c_{1}\right) - 4 c_{2}c_{3}, \nonumber \\
{\hat{c}}_{3} &\equiv& 2 c_{2}\left(3 c_{1} - c_{2}\right), \nonumber \\
{\hat{c}}_{4} &\equiv& 3 \epsilon c_{1}^2 C_{s} N_{0}^2 \left[ \eta_{0} /\left(3\omega + 4\right)) - 2 c_{2}^2\right].
\end{eqnarray}
A similar substitution into (\ref{FE(44)}) yields an expression for $\ddot{F}$, whose explicit form we omit, as $\ddot{F} = \ddot{F}(W, F, \dot{W}, \dot{F})$.  With these  $\ddot{W}$ and $\ddot{F}$  we find that (\ref{FE(11)}) reduces to (\ref{first-order equation for W and F}). As a result there are two independent equations, viz., (\ref{equation for ddot W, c1 neq 0}) and (\ref{first-order equation for W and F}), for the two unknowns $W$ and $F$. 

\paragraph{Ib:} For $c_{1} \neq 0$, $c_{2} = 0$ $(C \neq 0, \eta_{0} \neq 0)$, equation (\ref{equation for ddot W, c1 neq 0}) can be easily integrated.   The  general solution is 
$F = F_{0} \dot{W}^{- 1}   W^{c_{3}/c_{1}}$, where $F_{0}$ is a constant of integration.  Using this expression,  and (\ref{equation for Q}) with $c_{2} = 0$, we find that  both (\ref{FE(00)}) and (\ref{FE(44)}) generate the same second-order differential equation for $W$, namely,
\begin{equation}
\label{equation for W, case c2 = 0, c1 neq 0}
c_{1}\left(c_{1} - c_{3}\right) W \ddot{W}+ \frac{\dot{W}^2}{2}\left[c_{1}^2 \omega + \frac{4}{3}c_{3}^2 
- \frac{3 \epsilon c_{1}^2\left(c_{1}- c_{3}\right)^2 \left(C_{s} + C\right) N_{0}^2}{F_{0}^2 \left(3\omega + 4\right)}W^{2\left(c_{1} - c_{3}\right)/c_{1}}\right]   = 0. 
\end{equation}
On the other hand, (\ref{FE(11)}) becomes a third-order differential equation for $W$, which upon substitution of (\ref{equation for W, case c2 = 0, c1 neq 0}) into it reduces to the identity $0 = 0$.

$\bullet$ A particular solution to (\ref{equation for ddot W, c1 neq 0}), for $c_{2} = 0$,  is $W = $ constant, which leads to $Q =$ constant. In turn, (\ref{FE(00)}) and  (\ref{FE(44)}) become identical to each other, both requiring $c_{3} = c_{1}$, in agreement  with (\ref{equation for W, case c2 = 0, c1 neq 0}). With this requirement  (\ref{FE(11)}) yields $F \propto t$.   

\paragraph{II:} We now consider the case where $c_{1} = 0$ and $k \neq 0$ with $c_{2} \neq 0$ and $c_{2} = 0$ (cases IIa and IIb bellow).

\paragraph{IIa:} First we assume $c_{2} \neq 0$, in which case  from (\ref{first condition on F, Q and W}) we get $F = F_{0} W^{- c_{3}/c_{2}}$. Substituting this and (\ref{first integral for f}) into the wave equation (\ref{Dalambertian in 5D with y dependence}) we obtain\footnote{In general, when   $c_{2} \neq  0$ from (\ref{first condition on F, Q and W}) we get $F = F_{0} W^{- c_{3}/c_{2}} Q^{- 3 c_{1}/c_{2}}$ and  the wave equation (\ref{Dalambertian in 5D with y dependence}) yields 
\[
\label{equation for ddot W}
c_{2}Q W \ddot{W} - \dot{W} \left[c_{3} Q \dot{W} + 3 \left(c_{1} - c_{2}\right) W \dot{Q}\right] + \frac{\epsilon c_{2}^2 C_{s}N_{0}^2\left[c_{2} + 3 \left(c_{1} - c_{3}\right)\right]}{F_{0}^2 \left(3\omega + 4\right)} Q^{\left(6 c_{1} + c_{2}\right)/c_{2}} W^{2 \left(c_{2} + c_{3}\right)/c_{2}} = 0.
\]}

\begin{equation}
\label{equation for ddot W, c1 = 0, c2 neq 0}
c_{2}Q W \ddot{W} - \dot{W} \left(c_{3} Q \dot{W} - 3 c_{2} W \dot{Q}\right) + \frac{\epsilon c_{2}^2 C_{s}N_{0}^2\left(c_{2} 
- 3 c_{3}\right)}{F_{0}^2 \left(3\omega + 4\right)} Q W^{2 \left(c_{2} + c_{3}\right)/c_{2}} = 0.
\end{equation}
When $c_{2} \neq 0$ separability (\ref{reduced form}) requires $\eta_{0} C = 0$. Following the same steps as above, we obtain a first-order differential equation by substituting $\ddot{W}$ from (\ref{equation for ddot W, c1 = 0, c2 neq 0}) into (\ref{FE(00)}). Namely, 
\begin{equation}
\label{first-order equation for W and Q}
\left(c_{2}\omega + 2 c_{3}\right) \frac{\dot{W}^2}{W^2} + 6 \left(c_{3} - c_{2}\right)\frac{ \dot{W} \dot{Q}}{W Q} - 6 c_{2}\frac{\dot{Q}^2}{Q^2} - \frac{6 c_{2} N_{0}^2}{P_{0}^2}\frac{k}{Q^2}  - \frac{\epsilon c_{2} C_{s} N_{0}^2}{F_{0}^2\left(3\omega + 4\right)}\left[c_{2}^2 \left(2\omega + 3\right) - 2 c_{2}c_{3} + 3 c_{3}^2\right] W^{2 c_{3}/c_{2}} = 0. 
\end{equation}
Taking the time derivative of (\ref{first-order equation for W and Q})  and replacing into it the expression for $\ddot{W}$ obtained from (\ref{equation for ddot W, c1 = 0, c2 neq 0}), after some lengthly algebraic manipulations, we derive (\ref{FE(44)}), which now is a second-order differential equation for $Q$. On the other hand, (\ref{FE(11)}) reduces to (\ref{first-order equation for W and Q}) after we substitute $\ddot{W}$ and $\ddot{Q}$ into it. Thus, (\ref{equation for ddot W, c1 = 0, c2 neq 0})-(\ref{first-order equation for W and Q}) generate all the separable solutions with $c_{1} = 0$, $k \neq 0$ and $c_{2} \neq 0$.

\paragraph{IIb:} We now let $c_{2} = 0$ and assume $c_{3} \neq 0$. In this case one can generate solutions with $k \neq 0$ and $\eta_{0} C \neq 0$. From (\ref{first condition on F, Q and W}) it follows that $W = $ constant. Consequently, the wave equation (\ref{Dalambertian in 5D with y dependence}) is satisfied identically. From (\ref{FE(00)}) and  (\ref{FE(44)}) we obtain the differential equations that govern  $F$ and $Q$. Namely,  
\begin{equation}
\label{first-order equation for Q and F}
 \frac{\dot{Q}^2}{Q^2} + \frac{ \dot{F} \dot{Q}}{F Q}  + \left(\frac{N_{0}}{P_{0}}\right)^2 \frac{k}{Q^2} + \frac{\epsilon c_{3}^2 \left(C_{s} + C\right) N_{0}^2}{2\left(3\omega + 4\right) F^2} = 0,  
\end{equation}
and 
\begin{equation}
\label{equation for ddot Q, c1 = 0, c2 = 0}
Q \ddot{Q} + \dot{Q}^2 + k \left(\frac{N_{0}}{P_{0}}\right)^2  - \frac{\epsilon c_{3}^2 \left(C_{s} + C\right) N_{0}^2 Q^2}{2 \left(3\omega + 4\right) F^2 } = 0.
\end{equation}
Taking the time derivative of (\ref{first-order equation for Q and F}) and using (\ref{equation for ddot Q, c1 = 0, c2 = 0}) we obtain an expression for $\ddot{F}$, which we omit here, as $\ddot{F} = \ddot{Q}(F, Q, \dot{F}, \dot{W})$. Once again after some manipulations, one can verify that (\ref{FE(11)}) reduces to (\ref{first-order equation for Q and F}) after $\ddot{F}$ and $\ddot{Q}$ are substituted into it. 

\medskip

Thus, we have obtained the differential equations that generate all distinct cases of separable solutions.  They demonstrate, in a constructive way, that the FE are consistent with the assumption of separability. In the case where  $c_{1} = c_{2} = c_{3} = 0$ the metric functions become independent of the extra coordinate. For this case, the solutions to the FE as well as their cosmological applications have recently been discussed by the present author in \cite{LateAcceleration}.



\begin{thebibliography}{99}

\bibitem{Riess}{A.G. Riess {\it et al.,}  Supernova Search Team Collaboration, ``Observational Evidence from Supernovae for an Accelerating Universe and a Cosmological Constant", {\em Astron. J.}  {\bf 116} (1998) 1009 [arXiv:astro-ph/9805201].}
\bibitem{Perlmutter}{S. Perlmutter {\it et al.,} Supernova Cosmology Project Collaboration,``Measurements of Omega and Lambda from 42 High-Redshift Supernovae",   {\em Astrophys. J.} {\bf 517} (1999)
565 [arXiv:astro-ph/9812133].}

\bibitem{Liddle}{Andrew R Liddle, ``Acceleration of the Universe",  {\em New Astron.Rev.} {\bf 45} (2001) 235 [arXiv:astro-ph/0009491].}
\bibitem{Seto}{N. Seto, S. Kawamura and T. Nakamura, ``Possibility of Direct Measurement of the Acceleration of the Universe Using 0.1 Hz Band Laser Interferometer Gravitational Wave Antenna in Space",  {\em Phys.Rev.Lett.} {\bf 87} (2001) 221103 [arXiv:astro-ph/0108011].}

\bibitem{Knop}{R. A. Knop {\em et al.}, ``New Constraints on $\Omega_M$, $\Omega_\Lambda$, and w from an Independent Set of Eleven High-Redshift Supernovae Observed with HST",   {\em Astrophys. J.} {\bf 598} (2003)  102 [arXiv:astro-ph/0309368].}
\bibitem{Tonry}{J.L. Tonry {\it et al.}, ``Cosmological Results from High-z Supernovae", {\em Astrophys. J.} {\bf 594} (2003) 1 [arXiv:astro-ph/0305008v1].}
\bibitem{Lee}{A.T. Lee {\it et al.}, ``A High Spatial Resolution Analysis of the MAXIMA-1 Cosmic Microwave Background Anisotropy Data",  {\em Astrophys. J.} {\bf 561} (2001)  L1 [arXiv:astro-ph/0104459].}


\bibitem{Stompor}{R. Stompor {\it et al}, ``Cosmological implications of the MAXIMA-I high resolution Cosmic Microwave Background anisotropy measurement", {\em Astrophys. J.} {\bf 561} (2001)  L7 [arXiv:astro-ph/0105062].}
\bibitem{Halverson}{
N.W. Halverson {\it et al.}, ``DASI First Results: A Measurement of the Cosmic Microwave Background Angular Power Spectrum",  {\em Astrophys. J.} {\bf 568} (2002)  38 [arXiv:astro-ph/0104489].}

\bibitem{Netterfielf}{C.B. Netterfield {\it et al.}, ``A measurement by BOOMERANG of multiple peaks in the angular power spectrum of the cosmic microwave background",   {\em Astrophys. J.} {\bf 571} (2002) 604 [arXiv:astro-ph/0104460].}

\bibitem{Pryke}{C. Pryke, {\it et al.}, ``Cosmological Parameter Extraction from the First Season of Observations with DASI",    {\em Astrophys. J.} {\bf 568} (2002)  46 [arXiv:astro-ph/0104490].}
\bibitem{Spergel}{D.N. Spergel {\it et al.,} ``First Year Wilkinson Microwave Anisotropy Probe (WMAP) Observations: Determination of Cosmological Parameters", {\em Astrophys. J.Suppl.}  {\bf 148} (2003) 175 [arXiv:astro-ph/0302209].}
\bibitem{Sievers}{J. L. Sievers, {\it et al.,} ``Cosmological Parameters from Cosmic Background Imager Observations and Comparisons with BOOMERANG, DASI, and MAXIMA",   {\em Astrophys. J.}  {\bf 591} (2003)  599 [arXiv:astro-ph/0205387].}

\bibitem{Review}{E.J. Copeland, M. Sami and S. Tsujikawa, ``Dynamics of dark energy",  {\em Int. J. Mod. Phys.} {\bf D 15} (2006) [arXiv:hep-th/0603057].}


\bibitem{Peebles}{P. J. E. Peebles and  B. Ratra, ``The Cosmological Constant and Dark Energy", {\em Rev.Mod.Phys.} {\bf 75} (2003) 559  [arXiv:astro-ph/0207347].}
\bibitem{Padmanabhan0}{T. Padmanabhan, ``Cosmological Constant - the Weight of the Vacuum", {\em Phys.Rept.} {\bf 380} (2003)  235 [arXiv:hep-th/0212290].}

\bibitem{Zlatev}{I. Zlatev, L Wang and P. J. Steinhardt, ``Quintessence, Cosmic Coincidence, and the Cosmological Constant", {\em Phys. Rev. Lett.} {\bf 82} (1999) 896 [arXiv:astro-ph/9807002].}
\bibitem{Armendariz}{C. Armendariz, V. Mukhanov, P. J. Steinhardt, ``A Dynamical Solution to the Problem of a Small Cosmological Constant and Late-time Cosmic Acceleration", {\em Phys. Rev. Lett.} {\bf 85} (2000)  4438 [arXiv:astro-ph/0004134].}
\bibitem{Caldwell1}{R.R. Caldwell, R. Dave, P. J. Steinhardt, ``Cosmological Imprint of an Energy Component with General Equation of State", {\em Phys. Rev. Lett.} {\em 80} (1998) 1582 [arXiv:astro-ph/9708069].}
\bibitem{Deustua}{S.E. Deustua, R. Caldwell, P. Garnavich, L. Hui, A. Refregier, 
``Cosmological Parameters, Dark Energy and Large Scale Structure" [arXiv:astro-ph/0207293].}

\bibitem{DissipativeFluids}{L.P. Chimento, A.S. Jakubi and D. Pavon, ``Enlarged quintessence cosmology", {\em Phys.Rev.} {\bf D 62} (2000) 063508 [arXiv:astro-ph/0005070].}

\bibitem{Chaplygin gas 1}{M. C. Bento, O. Bertolami, A. A. Sen. ``Generalized Chaplygin gas, accelerated expansion and dark energy-matter unification", 	{\em Phys. Rev.}  {\bf D 66} (2002) 043507 [arXiv:gr-qc/0202064].}

\bibitem{Chaplygin gas 2}{Zong-Kuan Guo and Yuan-Zhong Zhang, ``Cosmology with a variable Chaplygin gas", {\em Phys. Lett.} {\bf B 645} (2007) 326 [arXiv:astro-ph/0506091].}

\bibitem{K-essence 1}{C. Armendariz-Picon, V. Mukhanov and P.J. Steinhardt, ``A dynamical solution to the problem of a small cosmological constant and late-time cosmic acceleration", {\em Phys. Rev. Lett.} {\bf 85} (2000) [arXiv:astro-ph/0004134].}

\bibitem{K-essence 2}{R. de Putter and E.V. Linder, ``Kinetic k-essence and Quintessence", {\em Astropart. Phys.} {\bf 28} (2007) 263 [arXiv:0705.0400].}

\bibitem{K-essence 3}{R.J. Scherrer, ``Purely kinetic k-essence as unified dark matter", {\em Phys. Rev. Lett.} {\bf 93} (2004) 011301 [arXiv:astro-ph/0402316].}

\bibitem{K-essence 4}{C. Bonvin, C. Caprini and R. Durrer, ``No-go theorem for k-essence dark energy", {\em Phys. Rev. Lett.} {\bf 97} (2006) 081303 [rXiv:astro-ph/0606584].}

\bibitem{domain walls}{L. Conversi, A. Melchiorri, L. Mersini and J. Silk, ``Are domain walls ruled out?", {\em Astropart. Phys.} {\bf 21} (2004) 443 [arXiv:astro-ph/0402529].}

\bibitem{Rachel}{R. Bean, ``A weak lensing detection of a deviation from General Relativity on cosmic scales" [arXiv:0909.3853]; ``Current constraints on the cosmic growth history" [arXiv:1002.4197v3]. }

\bibitem{Gilles}{Gilles Esposito-Farese, ``Summary of session A4 at the GRG $18$ conference: Alternative Theories of Gravity", {\em Class. Quantum Grav.} {\bf 25} (2008) 114017 [arXiv:0711.0332].}

\bibitem{Bertolami}{O. Bertolami and P.J. Martins, ``Nonminimal coupling and quintessence",  {\em Phys. Rev.} {\bf D 61} (2000) 064007 [arXiv:gr-qc/9910056]. }

\bibitem{Pavon}{ N. Banerjee and D. Pavon, ``Cosmic acceleration without quintessence", {\em Phys. Rev.} {\bf D 63} (2001) 043504 [arXiv:gr-qc/0012048].}

\bibitem{Sen}{S. Sen and A. A. Sen, ``Late time acceleration in Brans Dicke Cosmology", {\em Phys. Rev.} {\bf D 63} (2001) 124006 [	arXiv:gr-qc/0010092].}



\bibitem{Kim 1}{Hongsu Kim, ``Brans-Dicke Theory as a Unified Model for Dark Matter - Dark Energy", {\em Mon. Not. Roy. Astron. Soc.} {\bf 364} (2005) 813 [arXiv:astro-ph/0408577].}


\bibitem{Kim 2}{Hongsu Kim, ``Brans-Dicke Scalar Field as a Unique k-essence", 
    {\em Phys. Lett.} {\bf B 606} (2005) 223 [arXiv:astro-ph/0408154].}

\bibitem{Das 1}{S. Das and  N. Banerjee, ``An interacting scalar field and the recent cosmic acceleration" {\em Gen. Rel. Grav.} {\bf 38} (2006) 785 [arXiv:gr-qc/0507115].}

\bibitem{Chakraborty}{W. Chakraborty and U. Debnath, ``Role of Brans-Dicke Theory with or without self-interacting potential in cosmic acceleration", {\em Int. J. Theor. Phys.} {\bf 48} (2009) 232 [arXiv:0807.1776].}

\bibitem{Dicke}{C. Brans and R.H. Dicke, ``Mach's Principle and a Relativistic Theory of Gravitation", {\em Phys. Rev} {\bf 124} (1961) 925.}

\bibitem{Das 2}{S. Das and  N. Banerjee, ``Brans-Dicke Scalar Field as a Chameleon", {\em Phys. Rev.} {\bf D 78} (2008) 043512 [arXiv:0803.3936].}

\bibitem{Li}{Li-e Qiang, Yongge Ma, Muxin Han, Dan Yu, ``5-dimensional Brans-Dicke Theory and Cosmic Acceleration" {\em Phys. Rev.} {\bf D 71} (2005) 061501 [arXiv:gr-qc/0411066].}

\bibitem{Li 1}{Li-e Qiang, Yan Gong, Yongge Ma and Xuelei Chen, ``Cosmological Implications of 5-dimensional Brans-Dicke Theory", To appear in PLB [arXiv:0910.1885].}

\bibitem{Aguilar}{J. E. Madriz Aguilar, C. Romero and A. Barros, ``Modified Brans-Dicke theory of gravity from five-dimensional vacuum", {\em Gen. Rel. Grav.} {\bf 40} (2008) 117 [arXiv:0705.0548].}

\bibitem{WJPdeL}{P.S. Wesson and J. Ponce de Leon, ``Kaluza-Klein equations, Einstein's equations, and an effective energy-momentum tensor", {J. Math. Phys.} {\bf 33} (1992) 3883.}

\bibitem{equiv}{J. Ponce de Leon, ``Equivalence Between Space-Time-Matter and Brane-World Theories", {\em Mod. Phys. Lett.} {\bf A 16} (2001) 2291 [arXiv:gr-qc/0111011].}

\bibitem{BOOMERANG}{P. de Bernardis {\em et al}, ``A flat universe from high-resolution maps of the cosmic microwave background radiation" {\em Nature} {\bf 404}, 955(2000). }









 

 





\bibitem{well-known}{J. Ponce de Leon, ``Cosmological models in a Kaluza-Klein theory with variable rest mass", {\em Gen. Rel. Grav.} {\bf 20} (1988) 539.}

\bibitem{XtraSymmetry}{J. Ponce de Leon, ``Extra symmetry in the field equations in 5D with spatial spherical symmetry", {\em Class. Quantum Grav.} {\bf 23} (2006) 3043 [arXiv:gr-qc/0512067].}

\bibitem{reinventing}{J. Ponce de Leon, ``Reinventing spacetime on a dynamical hypersurface", {\em Mod. Phys. Lett.} {\bf A 21} (2006) 947 [arXiv:gr-qc/0511067].}

\bibitem{Tae Hoon}{T.H. Lee, ``Some power-law cosmological solutions derived from the $5D$ Brans-Dicke vacuum theory", {\em Class. Quantum Grav.} {\bf 26} (2009) 137001.}



\bibitem{Wendy}{W.L. Freedman and Michael S. Turner, ``Measuring and Understanding the Universe", {\em Rev. Mod. Phys.} {\bf 75} (2003) 1433 [	arXiv:astro-ph/0308418].}

\bibitem{La}{D. La and P.J. Steinhardt, ``Extended Inflationary Cosmology", {\em Phys. Rev. Lett.} {\bf 62} (1989) 376.}

\bibitem{Gaztanaga}{E. Gaztanaga and A. Lobo, ``Non-Linear gravitational growth of large scale structures inside and outside standard Cosmology", {\em Astrophys. J.} {\bf 548} (2001) 47 [arXiv:astro-ph/0003129].}

\bibitem{LateAcceleration}{J. Ponce de Leon, ``Late time cosmic acceleration from vacuum Brans-Dicke theory in $5D$", {\em Class. Quantum Grav.} {\bf 27} (2010) 095002 [arXiv:0912.1026].}













\end{thebibliography}
\end{document}